\documentclass[11pt,letterpaper]{article}
\usepackage{graphicx}
\usepackage{jcappub}
\usepackage{setspace}
\usepackage{longtable}
\usepackage{tabularx}
\usepackage{amsmath,amssymb,color,mathrsfs,verbatim,bbm,wasysym,pstricks,epsfig,colortbl}
\usepackage{slashed,mathtools,youngtab,xcolor,rotating, placeins, rotating}
\usepackage[utf8]{inputenc}

\definecolor{niceblue}{RGB}{255,43,43}
\definecolor{nicered}{rgb}{0.01, 0.337, 0.6}
\definecolor{nicegreen}{RGB}{46, 139, 87}
\definecolor{mypurple}{RGB}{200, 100, 200}
\definecolor{niceorange}{RGB}{238, 121, 66}
\definecolor{magentish}{RGB}{0,200,200}
\definecolor{darkergreen}{rgb}{0,0.6666666,0}
\definecolor{darkerpurple}{rgb}{0.3333333,0,0.3333333}

\usepackage{hyperref}
\usepackage{cleveref}

\title{Cosmological Constraints on Interacting Light Particles} 
\author[a]{Christopher Brust,}
\author[b, a]{Yanou Cui,}
\author[c, d]{and Kris Sigurdson}
\affiliation[a]{Perimeter Institute for Theoretical Physics, \\
31 Caroline Street N, Waterloo, ON N2L 2Y5, Canada}
\affiliation[b]{Department of Physics and Astronomy, University of California, Riverside, CA 92521, USA}
\affiliation[c]{Department of Physics and Astronomy, University of British Columbia, \\
6224 Agricultural Road, Vancouver, BC V6T 1Z1, Canada}
\affiliation[d]{School of Natural Sciences, Institute for Advanced Study, 1 Einstein Drive, Princeton, New Jersey 08540, USA}
\emailAdd{cbrust@perimeterinstitute.ca, yanou.cui@ucr.edu, krs@phas.ubc.ca}
\abstract{
Cosmological observations are becoming increasingly sensitive to the effects of light particles in the form of dark radiation (DR) at the time of recombination. The conventional observable of effective neutrino number, $N_{\rm eff}$, is insufficient for probing generic, interacting models of DR. In this work, we perform likelihood analyses which allow both free-streaming effective neutrinos (parametrized by $N_{\rm eff}$) and interacting effective neutrinos (parametrized by $N_{\rm fld}$). We motivate an alternative parametrization of DR in terms of $N_{\rm tot}$ (total effective number of neutrinos) and $f_{\rm fs}$ (the fraction of effective neutrinos which are free-streaming), which is less degenerate than using $N_{\rm eff}$ and $N_{\rm fld}$. Using the Planck 2015 likelihoods in conjunction with measurements of baryon acoustic oscillations (BAO), we find constraints on the total amount of beyond the Standard Model effective neutrinos (both free-streaming and interacting) of $\Delta N_{\rm tot} < 0.39$ at 2$\sigma$. In addition, we consider the possibility that this scenario alleviates the tensions between early-time and late-time cosmological observations, in particular the measurements of $\sigma_8$ (the amplitude of matter power fluctuations at 8$h^{-1}$ Mpc), finding a mild preference for interactions among light species. We further forecast the sensitivities of a variety of future experiments, including Advanced ACTPol (a representative CMB Stage-III experiment), CMB Stage-IV, and the Euclid satellite. This study is relevant for probing non-standard neutrino physics as well as a wide variety of new particle physics models beyond the Standard Model that involve dark radiation.}


\begin{document}

\maketitle
\flushbottom

\newpage

\section{Introduction}

The era of high-precision cosmology which we are entering provides a variety of powerful observational tools to probe new particle physics beyond the Standard Model (BSM). In particular, cosmological observables can be uniquely sensitive to new particles with very weak or even only gravitational interactions with the SM which are beyond the reach of any terrestrial particle physics experiments. Through the previous decade into the next, we are entering an era of unprecedented experimental precision with which we will be able to refine and further test the paradigm of $\Lambda$CDM ($\Lambda$ Cold Dark Matter) cosmology. A particular class of BSM physics models that will be tested with significantly improved sensitivity are those involving new free-streaming sub-eV mass particles, which may reveal themselves through effective neutrino number, $N_{\rm eff}$ \cite{Jungman:1995bz, Dolgov:2002wy, Steigman:2001px, Lopez:1998aq}, in the Cosmic Microwave Background (CMB). In the form of relativistic dark radiation (DR), such light species can contribute an appreciable fraction of the total cosmic energy density during the epoch in which anisotropies are frozen in, which nearly coincides with matter-radiation equality, and thus DR is observable from the evolution of cosmological perturbations. The presence of dark radiation in our universe may naturally descend from symmetries in the underlying quantum field theory; for example, a shift symmetry leads to a spin-0 Nambu-Goldstone boson, a chiral symmetry can lead to a spin-1/2 fermion, and a gauge symmetry leads to a spin-1 vector boson \cite{Nakayama:2010vs, Fischler:2010xz, Chacko:2015noa}. 
Investigations related to the $N_{\rm eff}$ observable may also reveal non-standard neutrino physics as well as unconventional cosmic thermal histories that enhance or dilute the energy density in SM neutrinos. A recent analysis from the Planck collaboration \cite{Ade:2015xua} reports a measurement of $N_{\rm eff}=3.15\pm 0.23$ \footnote{This quoted result is based on Planck+BAO data, without including Planck's high-$\ell$ polarization data.}, which has constrained many new physics possibilities that were previously comfortably allowed by BBN and WMAP measurements. Nevertheless, new light degrees of freedom which decoupled from the SM at temperatures comparable to the temperature of the QCD phase transition or higher are still generally compatible with Planck constraints \cite{Brust:2013xpv}. However, the next-generation experiments of the coming decade, culminating in a CMB Stage-IV (CMB-S4) experiment \cite{Stebor:2016hgt, Abazajian:2016yjj}, herald another order-of-magnitude improvement in the measurement of $N_{\rm eff}$, down to the level of $\sigma(N_{\rm eff}) \sim 0.015-0.03$ \cite{Wu:2014hta, Abazajian:2016yjj, Errard:2015cxa}. Such level of precision is remarkable as it is comparable to a theoretical benchmark of $\Delta N_{\rm eff}\geq0.027$ \cite{Brust:2013xpv, Chacko:2015noa, Adshead:2016xxj}, corresponding to just a single real scalar which decoupled from the SM at temperatures between the weak scale and the cutoff of the SM effective field theory. 

Nevertheless, most of the earlier studies on constraining new light species in the CMB, including the Planck report, implicitly assume that these particles would be very weakly interacting and thus free-streaming at the time of recombination, just like the SM neutrinos. At the same time, the presence of effectively \textit{interacting} light species in the CMB is also well-motivated from a wide variety of BSM particle physics models. In addition, the possibility that the SM neutrinos have at least some BSM interactions has been considered in \cite{Chacko:2003dt, Friedland:2007vv, Cyr-Racine:2013jua}, which may be motivated by certain neutrino mass generation mechanisms. Recently, there has been an increased interest in the scenario of hidden dark sectors inspired by both dark matter (DM) physics (for example, \cite{Kolb:1985bf, Hodges:1993yb, Feng:2008mu, Ackerman:mha, Kaplan:2009de,
  Fan:2013yva,Foot:2014uba}) and certain solutions to the electroweak hierarchy problem (such as the twin Higgs model \cite{Chacko:2005pe, Chacko:2016hvu, Craig:2016lyx}). This interest has motivated considerations of generic BSM dark radiation species that are interacting in the CMB \cite{Jeong:2013eza, Chacko:2015noa, Buen-Abad:2015ova}. Possible particle physics identities of such interacting DR include e.g. dark gluons with non-abelian gauge self-interactions, or a plasma of light dark chiral fermions interacting with massless dark photons. Furthermore, it has been recently demonstrated that the presence of strongly interacting DR, along with (partially) interacting DM, may alleviate the notable ``anomalies'' associated with late-time observables: the $\sim3\sigma$ tensions between direct measurements of $H_0$ (the Hubble expansion rate today) and measurements of Large Scale Structure (LSS), in particular $\sigma_8$ (the linear theory amplitude of matter power fluctuations at 8 $h^{-1}$ Mpc), and their inferred values from fitting early-time CMB data with the $\Lambda$CDM paradigm \cite{Lesgourgues:2015wza, Chacko:2016kgg}.

Are there observable differences between such interacting DR and neutrino-like free-streaming radiation? To first order, both contribute to the total background radiation energy density in the same manner. Nevertheless, there are subleading, more subtle effects that can distinguish the two. It was first discussed in \cite{Bashinsky:2003tk} that free-streaming species induce a universal phase shift in the high-$\ell$ peaks of CMB anisotropy spectrum due to their super-(sound)horizon propagation. \cite{Bashinsky:2003tk} also discusses the enhancement of matter fluctuations due to free-streaming neutrinos. Shortly afterwards, \cite{Weinberg:2003ur} pointed out a damping effect on CMB tensor modes caused by free-streaming species, similar to effects on scalar modes discussed earlier in \cite{Hu:1995en, Bashinsky:2003tk}. Recently, through analytic studies, it was pointed out in \cite{Chacko:2015noa} that interacting DR may induce non-trivial effects on these more subtle observables as well, yet in a different or opposite way relative to that by free-streaming species; for example, the effects on the magnitude and direction of the phase shift. It is only as of the recent Planck results that experiments started to be sensitive to the phase shift in the high-$\ell$ peaks \cite{Follin:2015hya, Baumann:2015rya}. However, the conventional one-parameter fit template based on $N_{\rm eff}$ implicitly assumes all possible light species are free-streaming. Due to the aforementioned observable differences in free-streaming vs. interacting DR, this approach may be far from optimal for either imposing a reliable constraint on generic BSM DR, or disentangling new physics from experimental data. It was proposed in \cite{Chacko:2015noa} that a two-parameter fit incorporating both types of DR is a more inclusive approach. Around that time, similar considerations triggered the work in \cite{Baumann:2015rya}, which along with presenting other related physics insights performed such a dedicated two-parameter analysis based on Planck data and a projection for CMB-S4. 

In this work, we conducted a more comprehensive study of the cosmological constraints on generic DR, considering data from a variety of present and future experiments, and allowing both $N_{\rm eff}$ and $N_{\rm fld}$ to float, where $N_{\rm fld}$ parametrizes the total number of ``fluid-like'', or interacting, effective number of neutrinos. Our work differs and extends in several regards with respect to the previous studies in \cite{Baumann:2015rya}. First, we choose a different parametrization of $\Lambda$CDM as well as imposing different priors on various parameters in our likelihood analyses, to be described further below, which offers modestly stronger constraints on the present-day constraints of $N_{\rm fld}$. Second, we extend the results of \cite{Baumann:2015rya} by considering additional cosmological measurements as well, in particular focusing on the recently recognized tensions between early-time and late-time measurements, finding mild evidence to support a scenario with non-vanishing $N_{\rm fld}$, which alleviates the tension with measurements of $\sigma_8$ arising from observations of large scale structure. Third, in addition to forecasting for a CMB Stage-IV experiment, we also forecast experiments whose results will be available sooner (in particular, Advanced ACTPol as a representative for the CMB Stage-III experiments, as well as the Euclid cosmic shear survey), in order to obtain a road map of our understanding of new light degrees of freedom over time. Finally, we use our results to motivate a new parametrization of new light degrees of freedom in our universe, $N_{\rm tot}$  (the sum of $N_{\rm eff}$ and $N_{\rm fld}$) and $f_{\rm fs}\equiv N_{\rm fld}/N_{\rm tot}$ (the fraction of effective neutrinos that are free-streaming), which have more independent, manifest physical meanings. This new parametrization alleviates the modest degeneracy between $N_{\rm eff}$ and $N_{\rm fld}$, and provides a tighter effective constraint on the allowed total light degrees of freedom.

We begin in section \ref{sec:likelihoods} by describing the details of our likelihood analyses, as well as defining the various likelihoods that we will use. Then, in section \ref{sec:results}, we describe our findings for the present-day results and future forecasts of these analyses and offer some interpretation of the results. Finally, in section \ref{sec:conclusions}, we conclude and describe future directions to pursue. In the Appendix we show a full triangle plot demonstrating the correlations between various cosmological parameters, upon digital zoom-in.

\section{Likelihood Analyses}
\label{sec:likelihoods}

We undertake several likelihood analyses of the scenario of having extra both interacting as well as free-streaming new light degrees of freedom at the time of recombination, parametrized in terms of an effective number of free-streaming neutrinos $N_{\rm eff}$ and an effective number of interacting\footnote{Generally these new light species may be either self-interacting (even possibly via higher-dimensional operators) or interact with other particles such as another type of DR or a component of dark matter \cite{Buen-Abad:2015ova, Lesgourgues:2015wza}. This work focuses on self-interaction and interaction among different types of DR. Interaction with dark matter would introduce a drag force on dark matter which is not included in our Boltzmann code.} neutrinos $N_{\rm fld}$, where our definition of ``interacting'' is that the interaction rate $\Gamma$ must be much larger than the Hubble expansion rate at the time of recombination which allows treating the dark radiation as fluid. To do this, we compare to cosmological likelihoods using a modified version of the Boltzmann code \texttt{CLASS v2.4.5} \cite{Blas:2011rf} in conjunction with the Monte Carlo code \texttt{MontePython v2.2.2} \cite{Audren:2012wb}. The modifications to CLASS implemented interacting particles as a perfectly coupled fluid. This models interacting dark radiation well as long as the fluid is tightly coupled; in other words, its interaction rate $\Gamma$ is much larger than $H$ throughout the epoch of recombination. The case where fluid self-interactions become inefficient during recombination ($\Gamma \sim H$) is more nuanced and we do not treat it here.

We compare primarily to the Planck 2015 release of cosmological likelihoods \cite{Aghanim:2015xee}, but in addition, we compare to other published cosmological measurements as well. We list the precise collection of data that we use later in subsections \ref{sec:presentlikelihoods} and \ref{sec:mocklikelihoods}. In addition, in 2016, Planck has released intermediate results \cite{Adam:2016hgk} using instead a cross-spectra-based approach to the low-$\ell$ polarization likelihood \cite{Mangilli:2015xya}, which they dubbed the ``Lollipop'' likelihood. 
They found that using Lollipop in conjunction with the Planck both low- and high-$\ell$ TT likelihoods lowered the posterior on the Thomson scattering optical depth $\tau$ from $0.078\pm 0.019$ to $0.058 \pm 0.012$, and the new result was stable upon further adding a lensing likelihood. It is interesting to determine whether this lower optical depth might have any effect on the parameters in the DR sector. We therefore study the constraints on new light degrees of freedom in the presence of a Gaussian prior on $\tau$, in lieu of low-$\ell$ polarization data from Planck. This serves as a mockup of the full Lollipop likelihood. The inclusion of a mockup of the Lollipop likelihood is an update to the results presented in the earlier related work in \cite{Baumann:2015rya}, in addition to the additional cosmological data we consider \footnote{At the time of the preparation of this draft, we found somewhat stronger present-day constraints between our analysis and theirs, which we have understood to be due to two driving factors. First, we impose Gaussian priors on nuisance parameters, as suggested by Planck. Second, we impose flat priors on the parameters $\ln(10^{10}A_s)$ and $\theta_s$ (as Planck does) rather than $A_s$ and $H_0$, as these parameters are less degenerate with $N_{\rm eff}$ and $N_{\rm fld}$. Although we have not done so, it would be interesting to approach the problem with a frequentist approach rather than a Bayesian one to eliminate the dependence on priors, and compare those constraints with the ones obtained here.}.

Finally, we also do a full likelihood forecast for various upcoming experiments, including Advanced ACTPol, a Euclid cosmic shear survey, and a tentative CMB-Stage IV experiment. The Stage-IV forecasting was also performed in \cite{Baumann:2015rya}; the main difference\footnote{Another difference is our combining with the Planck likelihood so as to extend our survey down to the low-$\ell$ regime; doing so allows us to more accurately consider potential correlations between our new DR parameters and $\Lambda$CDM parameters of interest, such as $H_0$ and $\sigma_8$.} in our analysis is a more conservative view of such an experiment, only working up to $\ell_{max}=3000$ instead of $5000$, and surveying 40\% of the sky instead of 75\%. We do not consider lensing reconstruction here in this initial treatment.

We now turn to the specific details of both our present-day and forecasted data sets.

\subsection{Present-Day Likelihoods}
\label{sec:presentlikelihoods}

We define the following combinations of likelihoods from the Planck 2015 release:

\begin{itemize}

\item {\bf ``Planck T''} includes the following likelihoods:
\begin{itemize}
\item High-$\ell$ TT from \texttt{hi\_l/plik/plik\_dx11dr2\_HM\_v18\_TT.clik}
\item Low-$\ell$ TEB from \texttt{low\_l/bflike/lowl\_SMW\_70\_dx11d\_2014\_10\_03\_v5c\_Ap.clik}
\item Lensing from \texttt{lensing/smica\_g30\_ftl\_full\_pp.clik\_lensing}
\end{itemize}

\item {\bf ``Planck P''} includes the following likelihoods:
\begin{itemize}
\item High-$\ell$ TT+TE+EE from \texttt{hi\_l/plik/plik\_dx11dr2\_HM\_v18\_TTTEEE.clik}
\item Low-$\ell$ TEB from \texttt{low\_l/bflike/lowl\_SMW\_70\_dx11d\_2014\_10\_03\_v5c\_Ap.clik}
\item Lensing from \texttt{lensing/smica\_g30\_ftl\_full\_pp.clik\_lensing}
\end{itemize}

\item {\bf ``BAO''} includes measurements of the following parameters at the following effective redshifts from the following surveys: 
\begin{itemize}
\item $r_{\mathrm{s}}/D_V$ at $z = 0.106$ from the 6dF galaxy survey \cite{Beutler:2011hx}
\item $D_V/r_{\mathrm{s}}$ at $z = 0.15$ from the SDSS main galaxy sample \cite{Ross:2014qpa}
\item $D_V/r_{\mathrm{s}}$ at $z = 0.32$ from BOSS-LOWZ \cite{Anderson:2013zyy}
\item $D_V/r_{\mathrm{s}}$ at $z = 0.57$ from BOSS-CMASS \cite{Anderson:2013zyy}
\end{itemize}

\item {\bf ``${\mathbf H_0}$''} includes a Gaussian prior on $H_0$ of $73.24\pm 1.74$ Mpc$^{-1}$km/s, from \cite{Riess:2016jrr}.

\item {\bf ``LSS''} includes a Gaussian priors on\footnote{Note that these measurements necessarily must assume some cosmology in order to produce quantitative results, and these were derived in the $\Lambda$CDM cosmology rather than the one we study in this paper. Nevertheless, we have verified that the introduction of new light degrees of freedom (evaluated at the best-fit point of our scan involving this likelihood below) does not significantly alter the matter power spectrum at the relevant scales, and so it is approximately correct to use these results as-is.}:
\begin{itemize}
\item $\sigma_8 (\Omega_m/0.27)^{0.46}$ of $0.774 \pm 0.040$ from the CFHTLenS weak lensing survey \cite{Heymans:2013fya}
\item $\sigma_8 (\Omega_m/0.27)^{0.30}$ of $0.782 \pm 0.010$ from the Planck SZ cluster mass function \cite{Ade:2013lmv}
\end{itemize}

Finally, in lieu of having the full cross-spectra likelihood, we use the following as a stand-in.

\item {\bf ``Lollipop''} includes the following likelihoods: 
\begin{itemize}
\item High-$\ell$ TT+TE+EE from \texttt{hi\_l/plik/plik\_dx11dr2\_HM\_v18\_TTTEEE.clik}
\item Low-$\ell$ TT only from \texttt{low\_l/commander/commander\_rc2\_v1.1\_l2\_29\_B.clik}
\item Lensing from \texttt{lensing/smica\_g30\_ftl\_full\_pp.clik\_lensing}
\item A gaussian prior on $\tau$ of $0.058\pm 0.012$, obtained from \cite{Adam:2016hgk}
\end{itemize}

\end{itemize}

\subsection{Mock Likelihoods}
\label{sec:mocklikelihoods}

\begin{itemize}

\item {\bf ``S3''} includes a mock likelihood created in \texttt{MontePython} based on the Planck P best-fit point for the {\it six}-parameter cosmological model, taken from \cite{Ade:2015xua}. We used the following parameters for the experiment, obtained from the specifications for the Stage-III five-channel Advanced ACTPol experiment \cite{Henderson:2015nzj} which is beginning to take data this year. These specifications are functionally equivalent to ones that could be obtained from the comparable SPT-3G experiment \cite{Benson:2014qhw}, Simons Array or POLARBEAR-2 \cite{Suzuki:2015zzg}.
\begin{itemize}
\item $\ell_{\mathrm{min}} = 30$
\item $\ell_{\mathrm{max}} = 2000$
\item $f_{\mathrm{sky}} = 0.4$
\item Angular resolution $=\{7.1', 4.8', 2.2', 1.4', 0.9'\}$
\item $\sigma_T = \frac{1}{\sqrt{2}}\sigma_P = \{80, 70, 8, 7, 25\} ' \mu K$
\end{itemize}

\item {\bf ``S4''} includes a mock likelihood also created in \texttt{MontePython} based on the same Planck E best-fit point. We used the following parameters for the tentative upcoming experiment, obtained from the CMB Stage-IV Science Book \cite{Abazajian:2016yjj}.
\begin{itemize}
\item $\ell_{\mathrm{min}} = 30$
\item $\ell_{\mathrm{max}} = 3000$
\item $f_{\mathrm{sky}} = 0.4$
\item Angular resolution $=1'$
\item $\sigma_T = \frac{1}{\sqrt{2}}\sigma_P = 1' \mu K$
\end{itemize}

\item {\bf ``Euclid''} includes a mock cosmic shear survey based on the upcoming Euclid experiment, scheduled to launch in 2020. The implementation of the mock survey was pulled from the Euclid Red Book \cite{Laureijs:2011gra}, and comes bundled with \texttt{MontePython}.
\end{itemize}

\section{Results}
\label{sec:results}

We ran Markov Chain Monte Carlo (MCMC) for a variety of combinations of the above likelihoods until the Gelman-Rubin convergence diagnostic $R-1 \lesssim 0.01$ for all chains. For the present-day constraints below, we also ran an otherwise identical scan but having fixed $N_{\rm eff} = 3.046$ and $N_{\rm fld}=0$ to reproduce $\Lambda$CDM, in order to report a $\Delta \chi^2\equiv \chi^2_{\Lambda\mathrm{CDM}+N_{\rm eff}+N_{\rm fld}} - \chi^2_{\Lambda\mathrm{CDM}}$. In all of the scans we discuss below, we take flat priors on the eight parameters $\Omega_b h^2$, $\Omega_{cdm} h^2$, $100 \theta_s$, $\ln(10^{10}A_s)$, $n_s$, $\tau$, $N_{\rm eff}$, and $N_{\rm fld}$. We of course impose a prior $N_{\rm fld}\geq 0$, and also impose a prior of $\tau\geq 0.04$ due to observations of the Gunn-Peterson effect (see e.g.  \cite{Caruana:2013qua}). Note that we impose the prior $N_{\rm eff} \geq 0$ rather than $N_{\rm eff} \geq 3.046$, allowing the possibility that the SM neutrinos have some BSM interactions, and the possibility of dilution of $N_{\rm eff}$ due to on-standard cosmology (e.g. by $N_{\rm fld}$ in our model, or by late-time entropy release after neutrino decoupling).

\subsection{Present-Day Constraints}

\begin{figure}[h]
\centering
\includegraphics[width=0.32\textwidth]{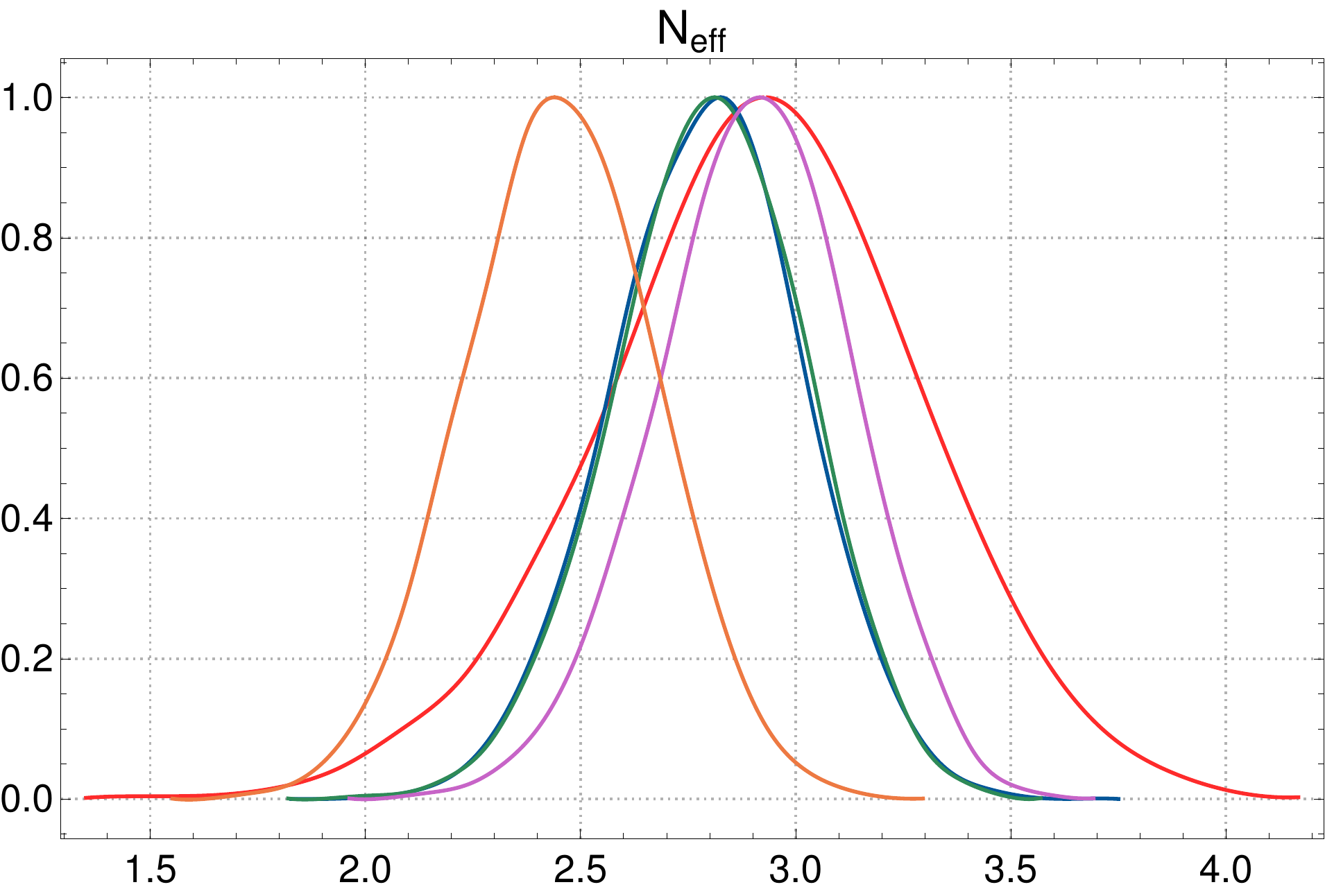}
\includegraphics[width=0.32\textwidth]{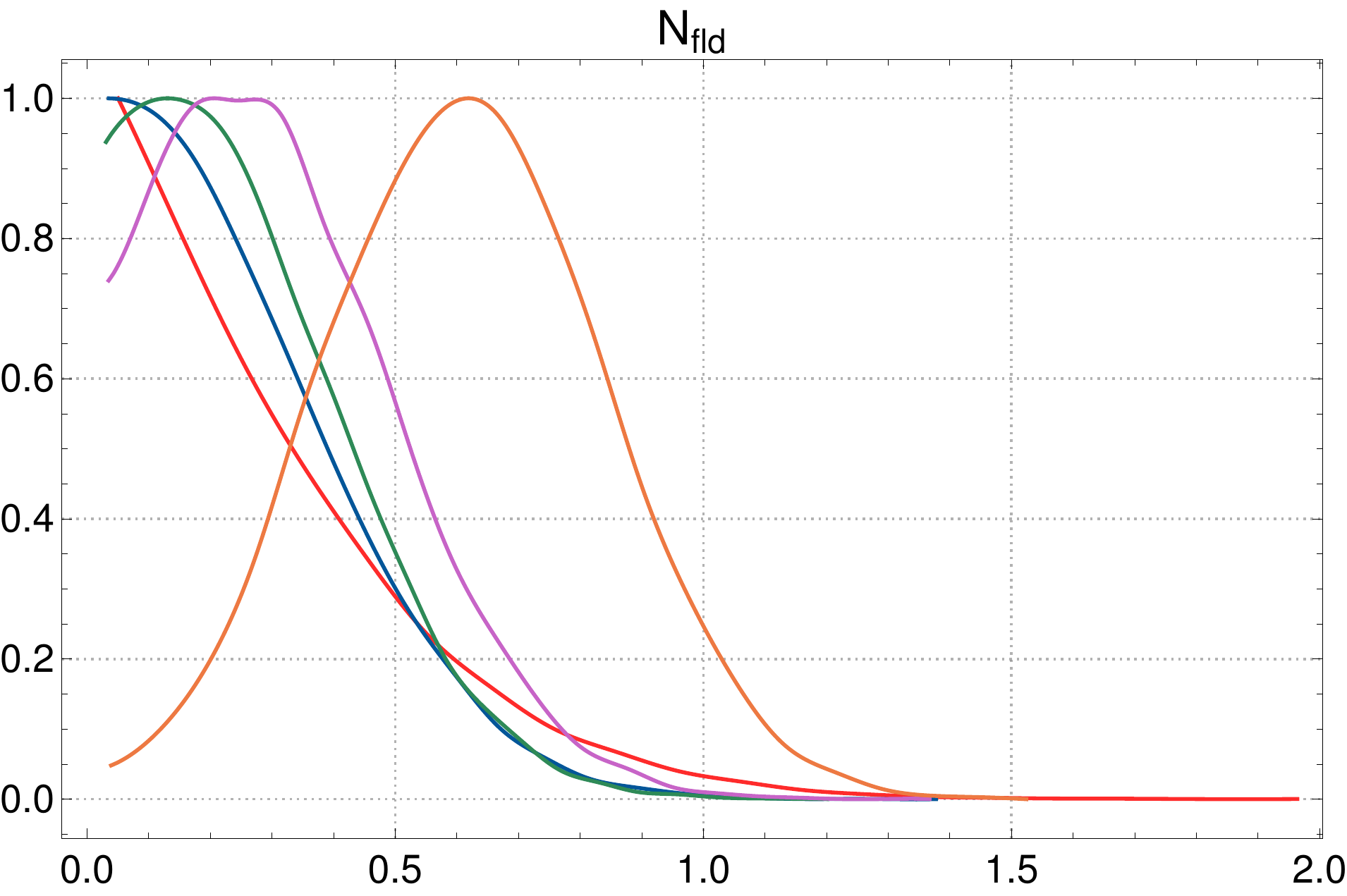}
\includegraphics[width=0.32\textwidth]{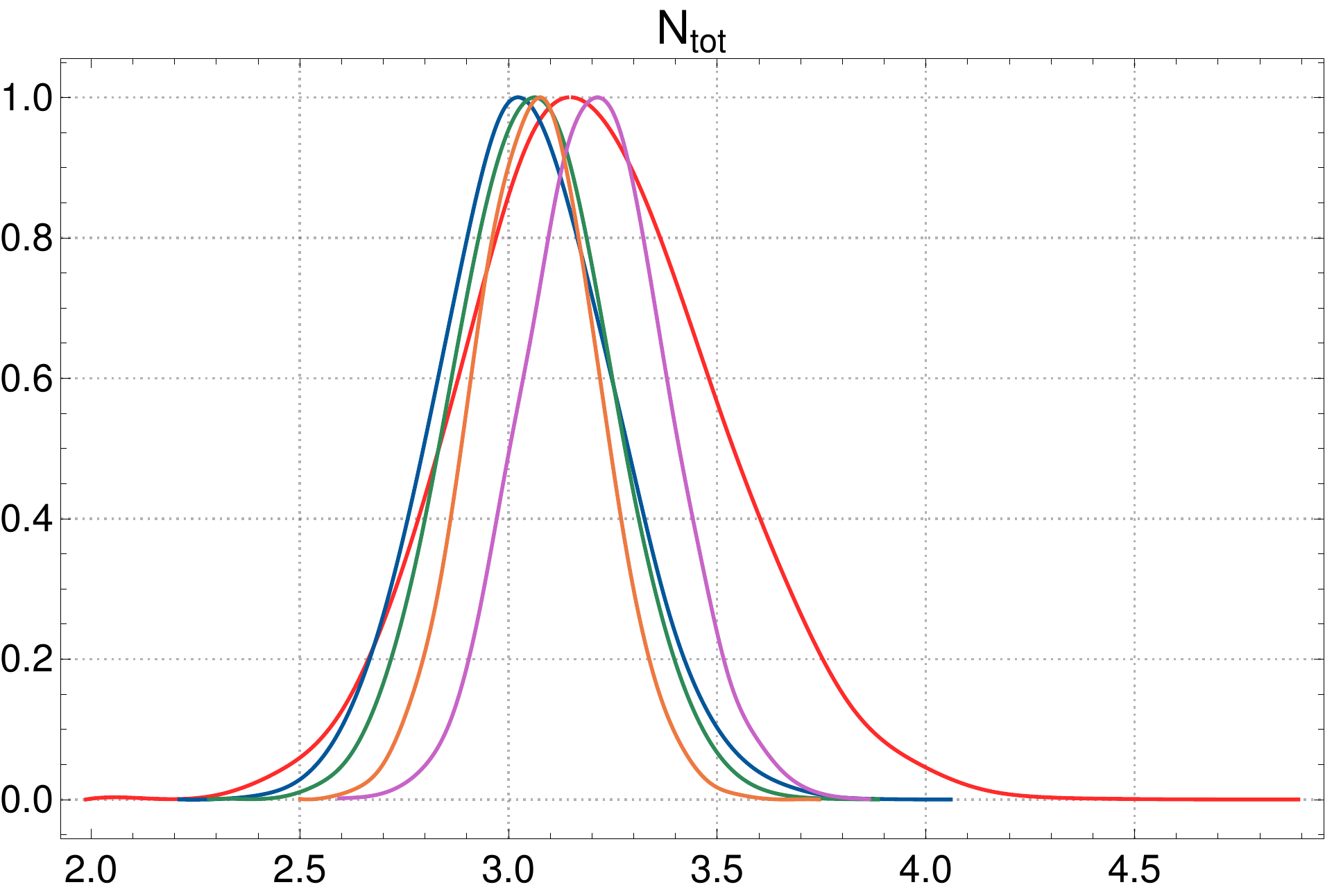}
\includegraphics[width=0.32\textwidth]{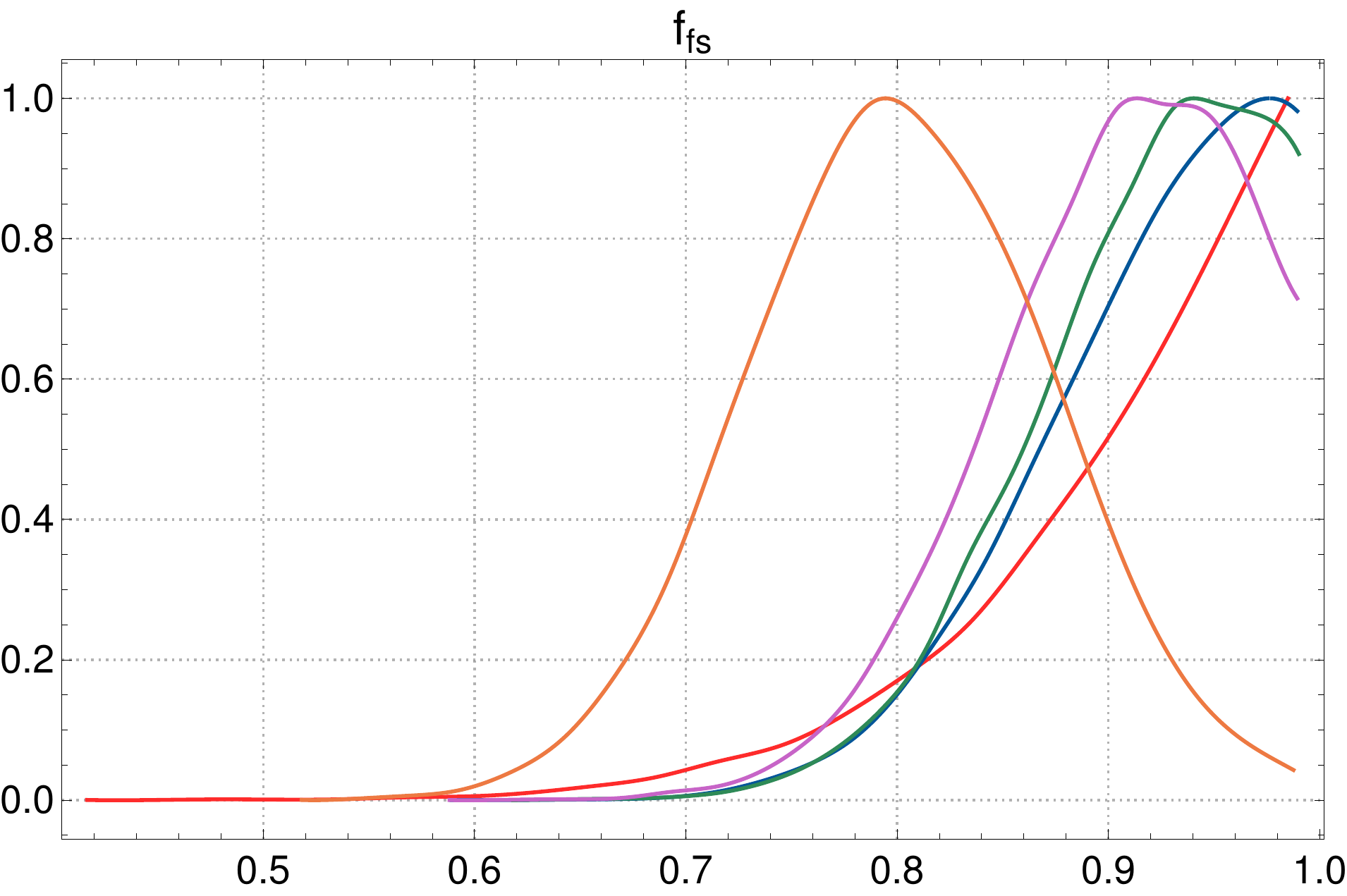}
\includegraphics[width=0.32\textwidth]{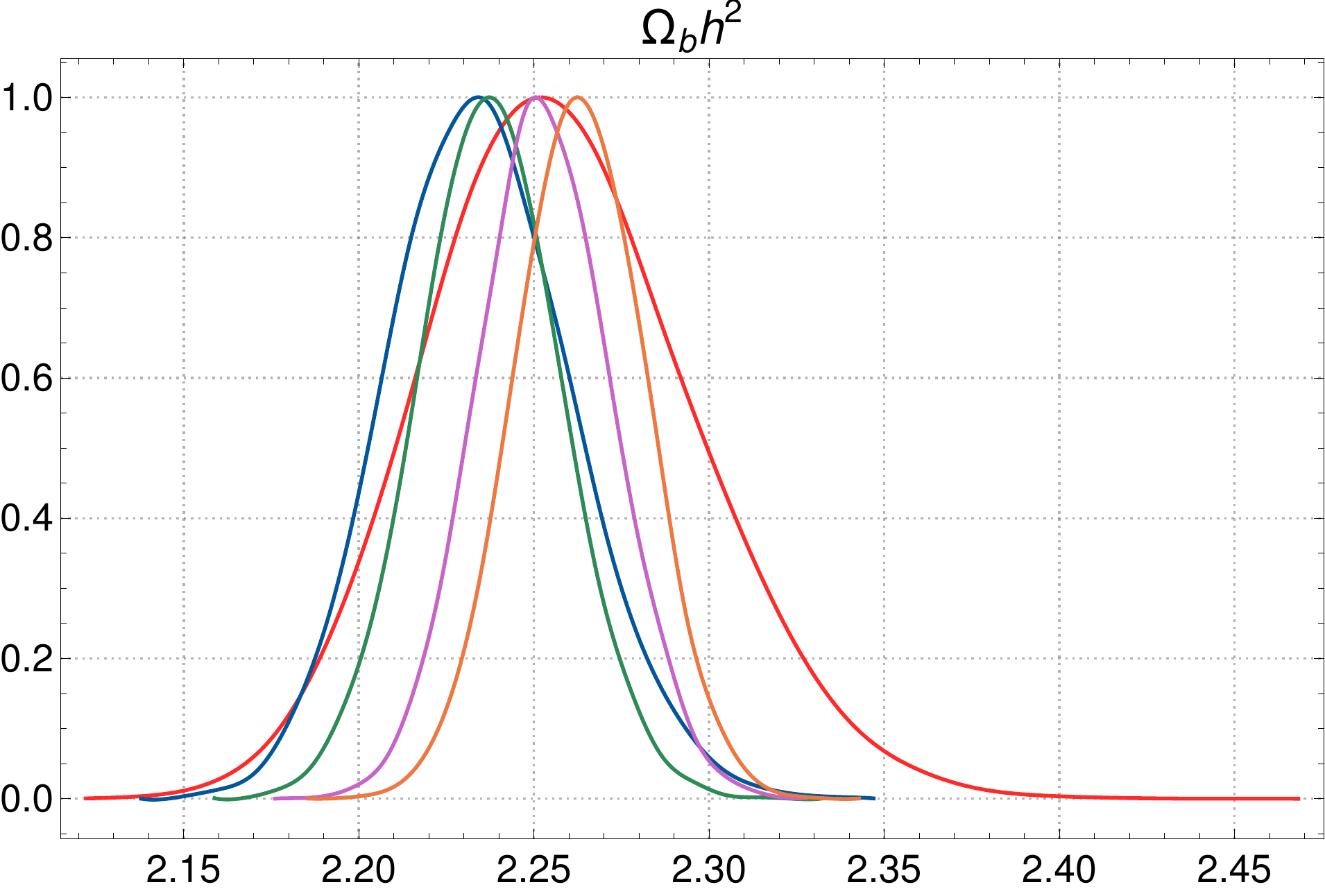}
\includegraphics[width=0.32\textwidth]{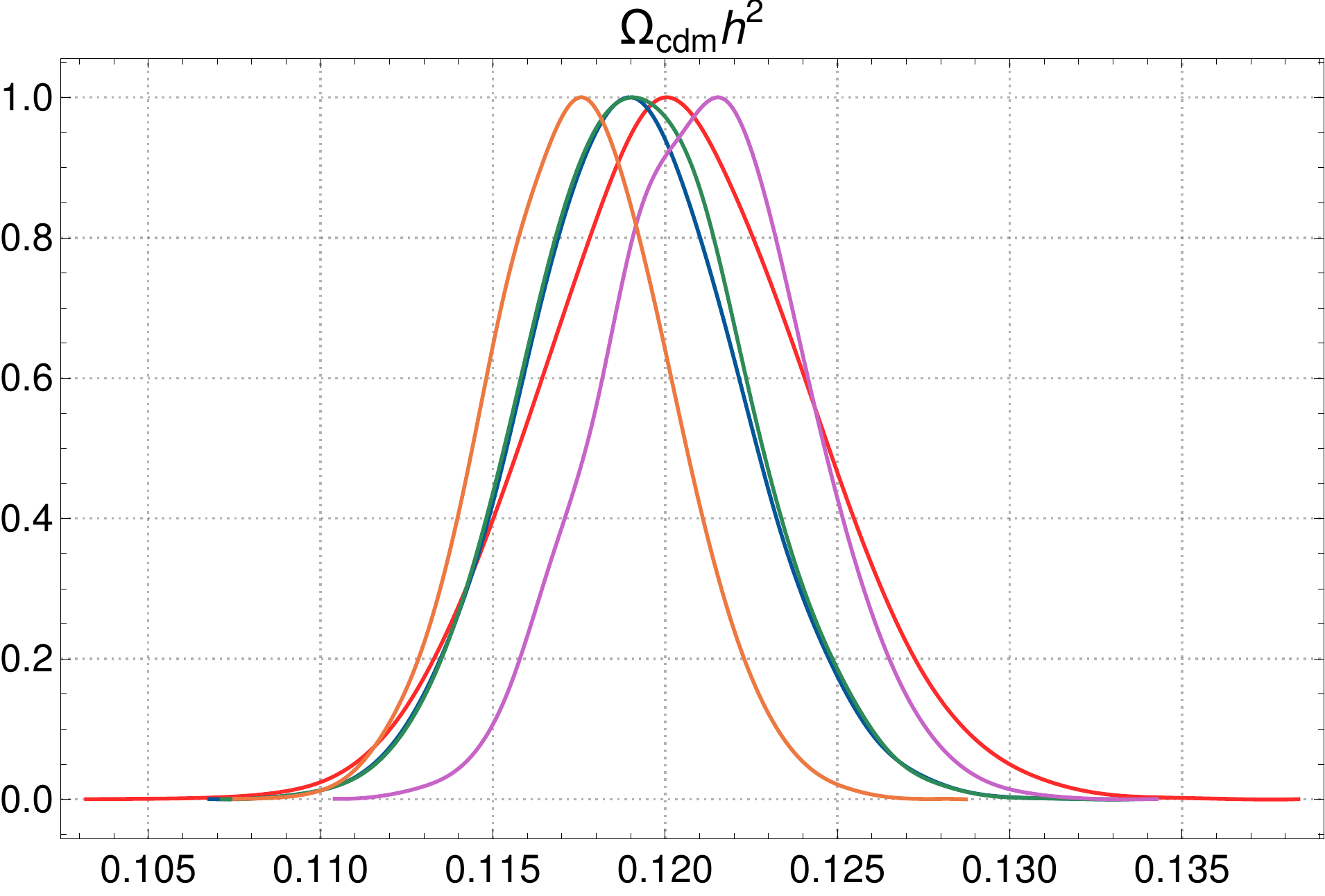}
\includegraphics[width=0.32\textwidth]{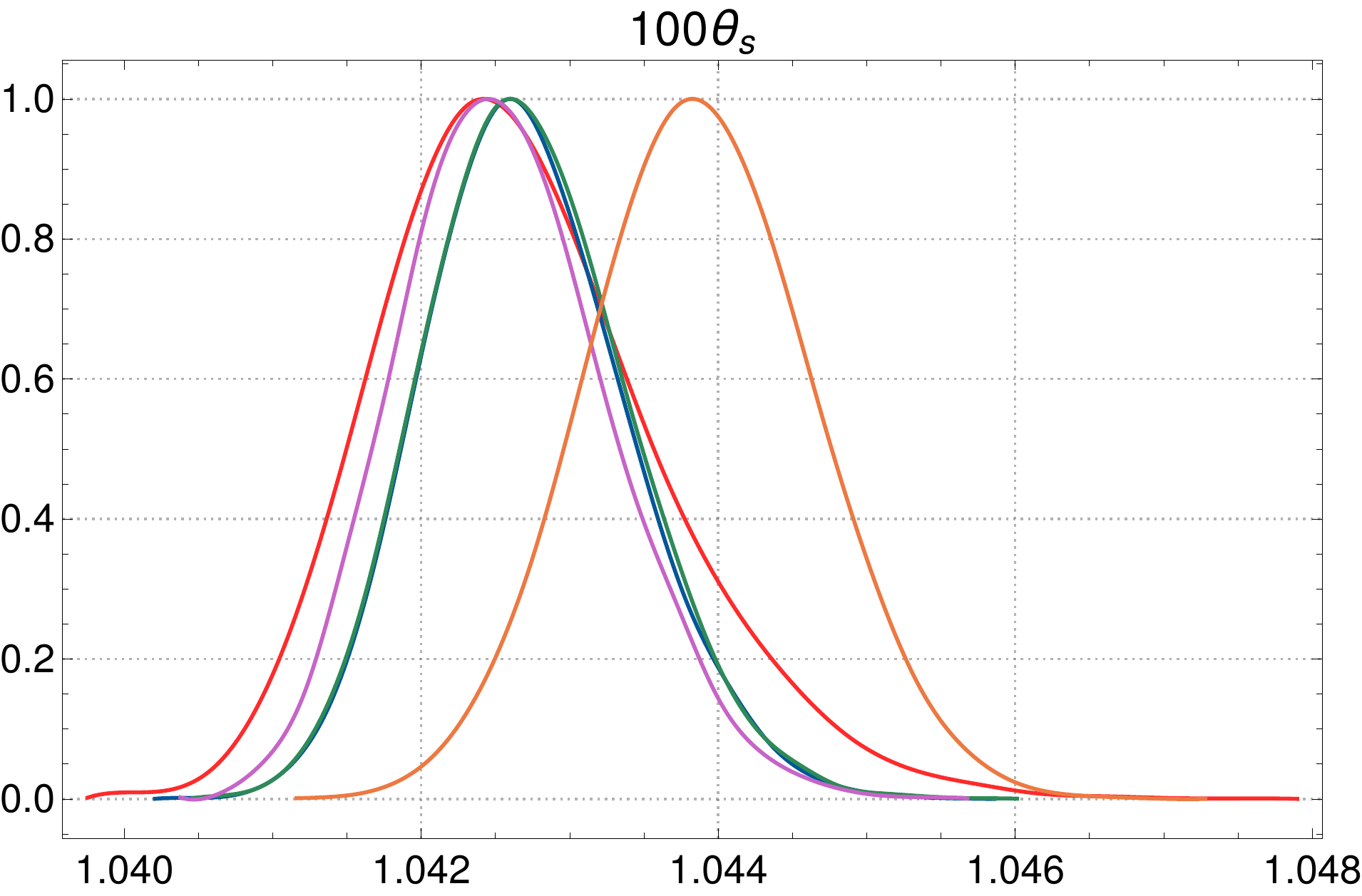}
\includegraphics[width=0.32\textwidth]{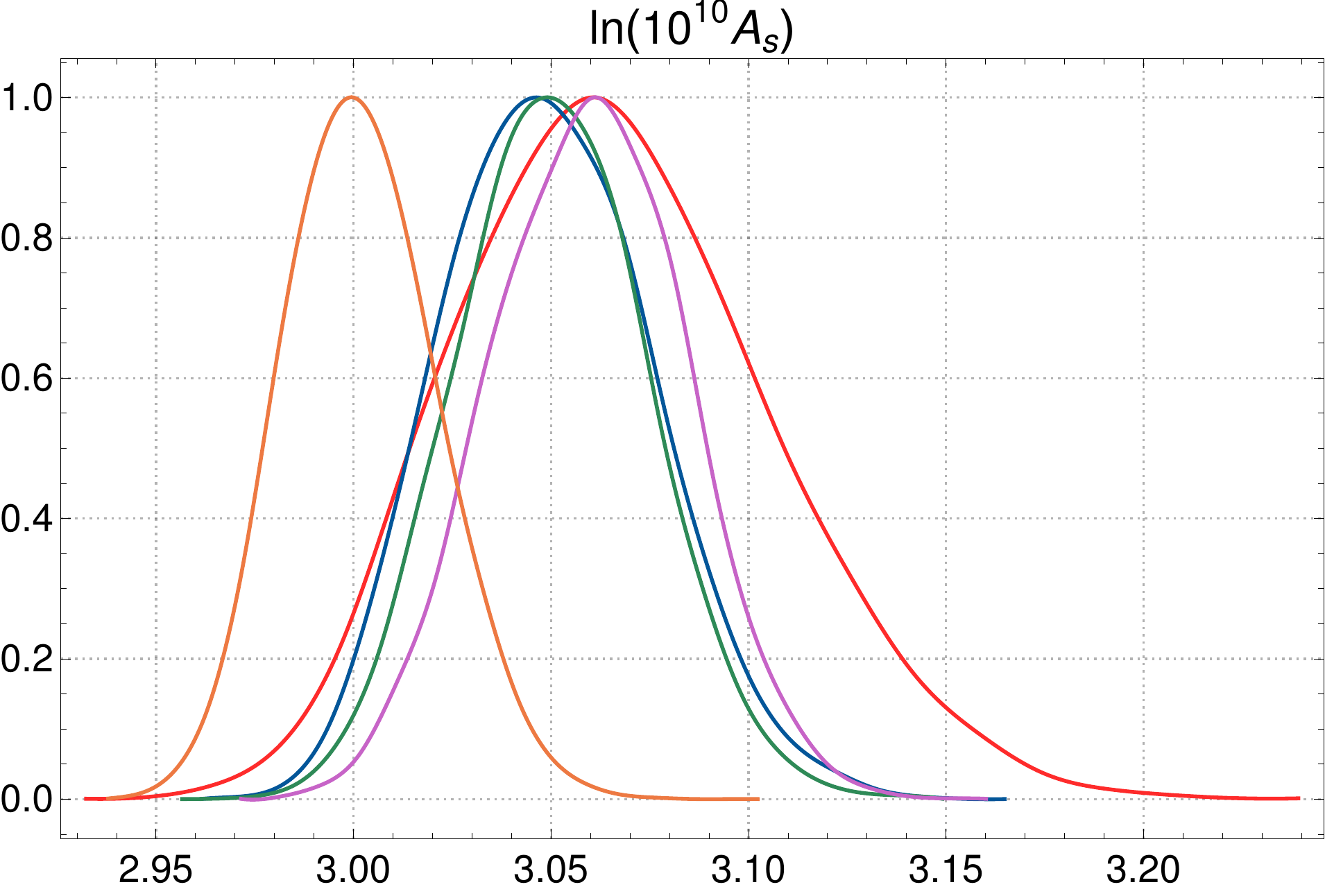}
\includegraphics[width=0.32\textwidth]{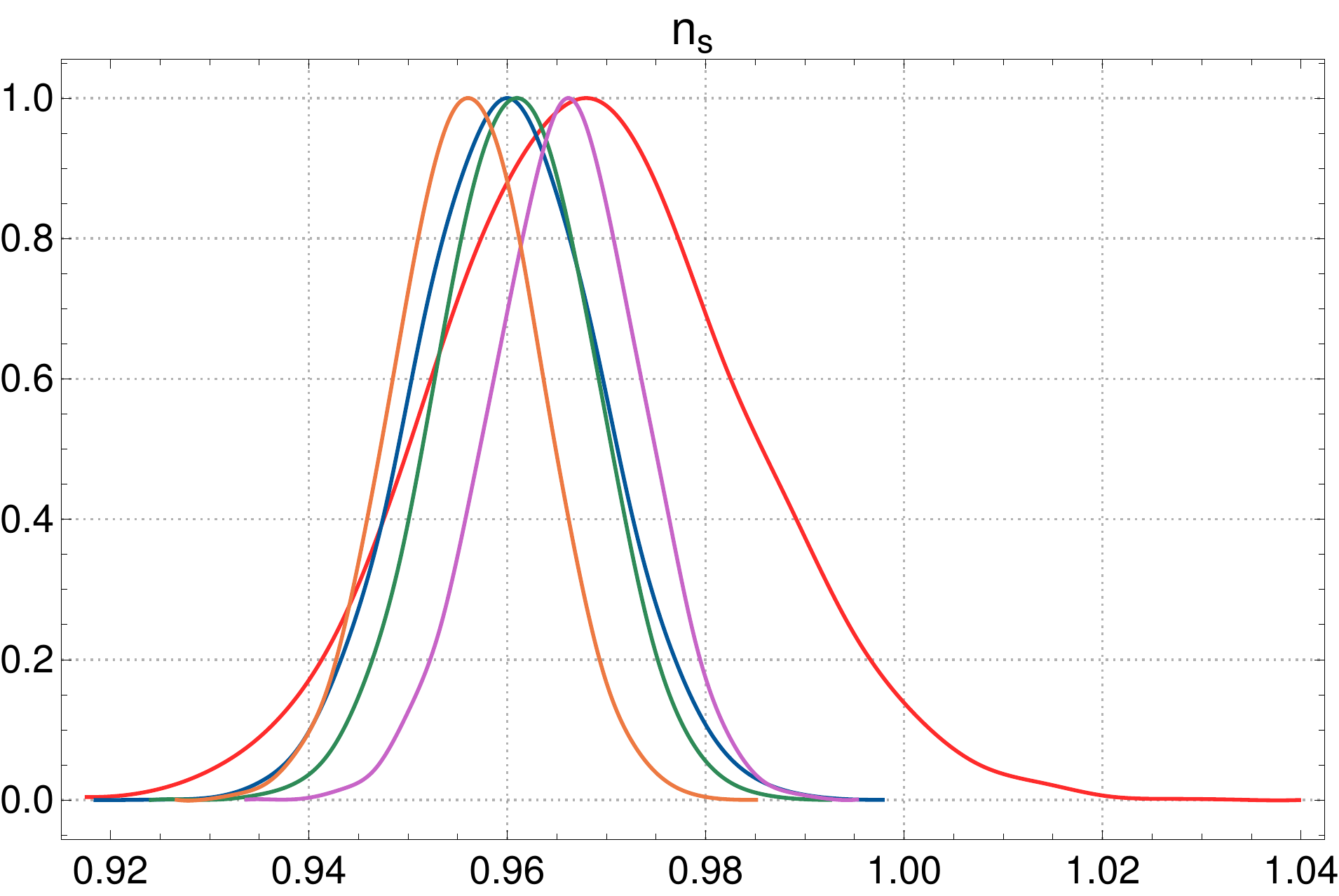}
\includegraphics[width=0.32\textwidth]{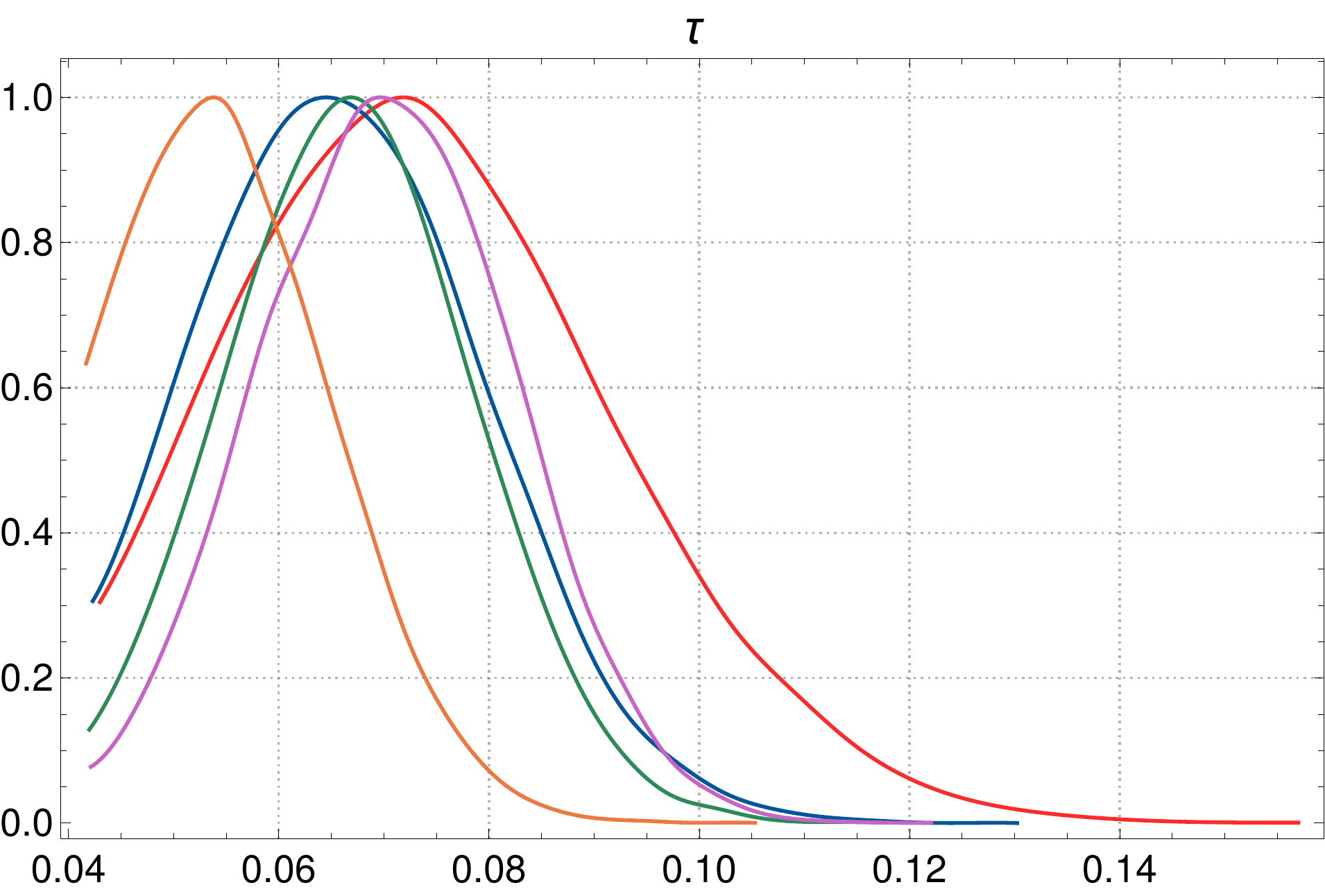}
\includegraphics[width=0.32\textwidth]{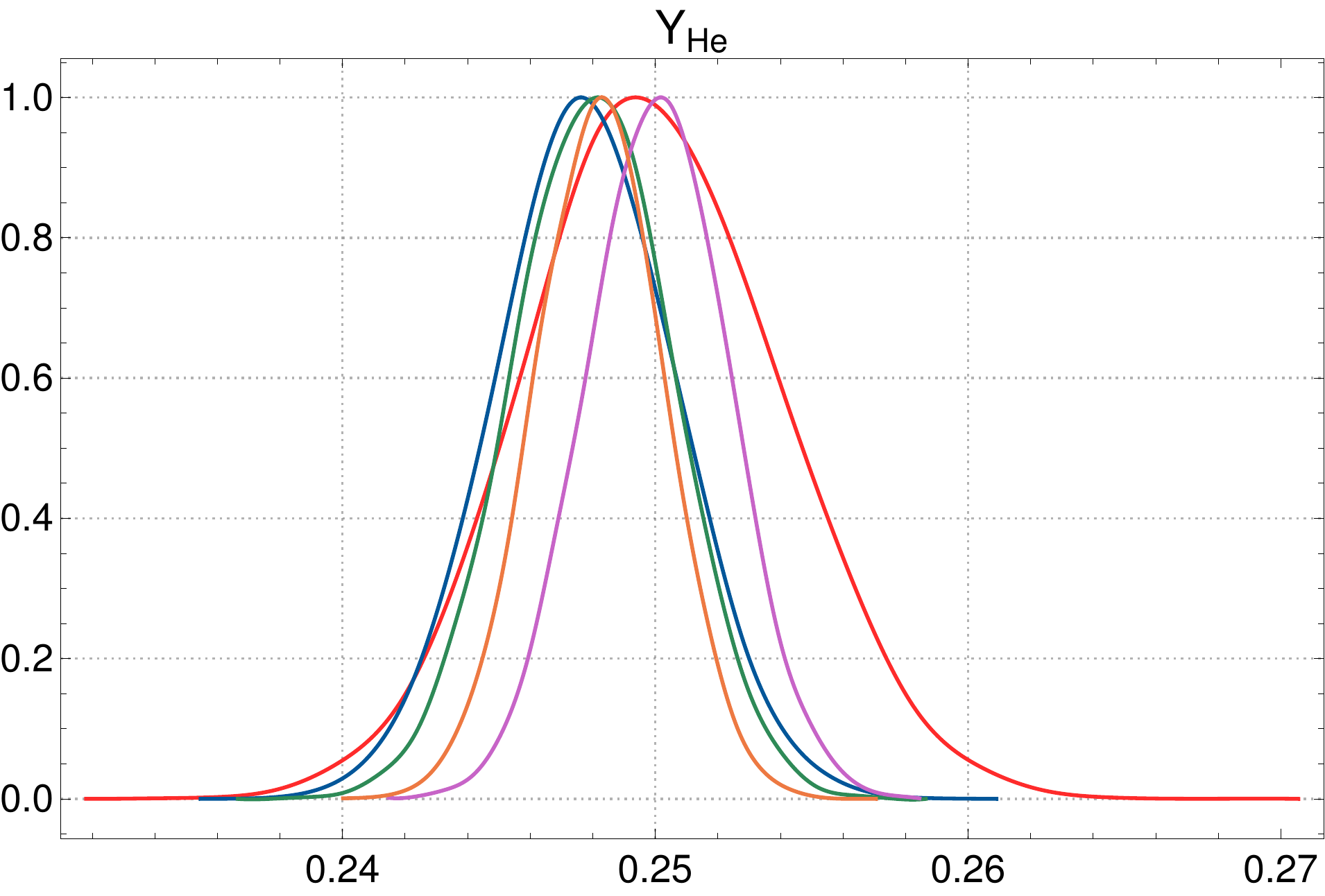}
\includegraphics[width=0.32\textwidth]{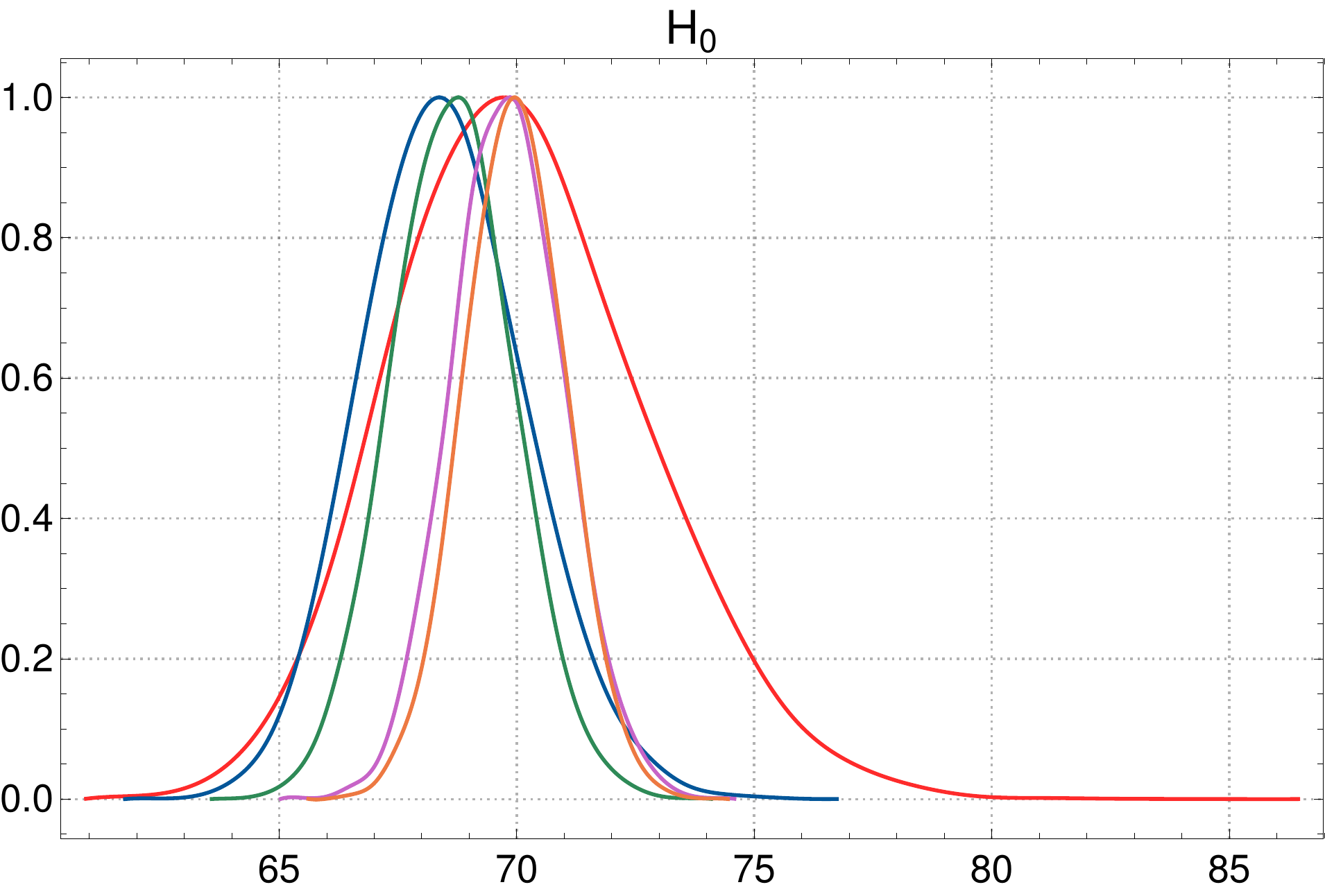}
\includegraphics[width=0.32\textwidth]{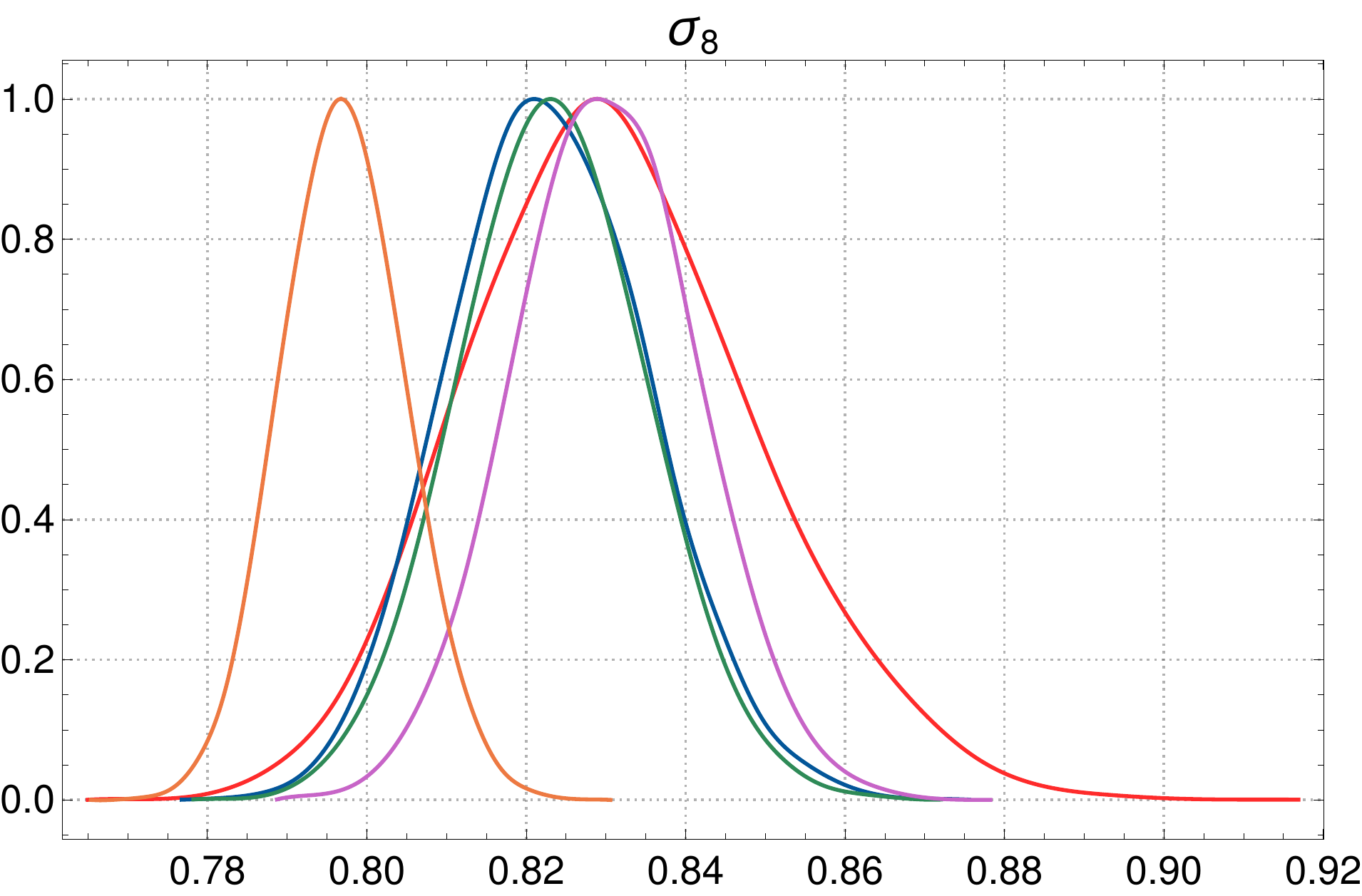}
\caption{\label{fig:1dpresentposteriors}Here we show the 1d posteriors for the above scans. The colors of the posterior indicate which scan they belong to; in the same order as above, {\color{niceblue} Planck T}, {\color{nicered} Planck P}, {\color{nicegreen} Planck P+BAO}, {\color{mypurple} Planck P+BAO+$H_0$}, and {\color{niceorange} Planck P+BAO+$H_0$+LSS}.}
\end{figure}

\begin{figure}[h]
\centering
\includegraphics[width=0.8\textwidth]{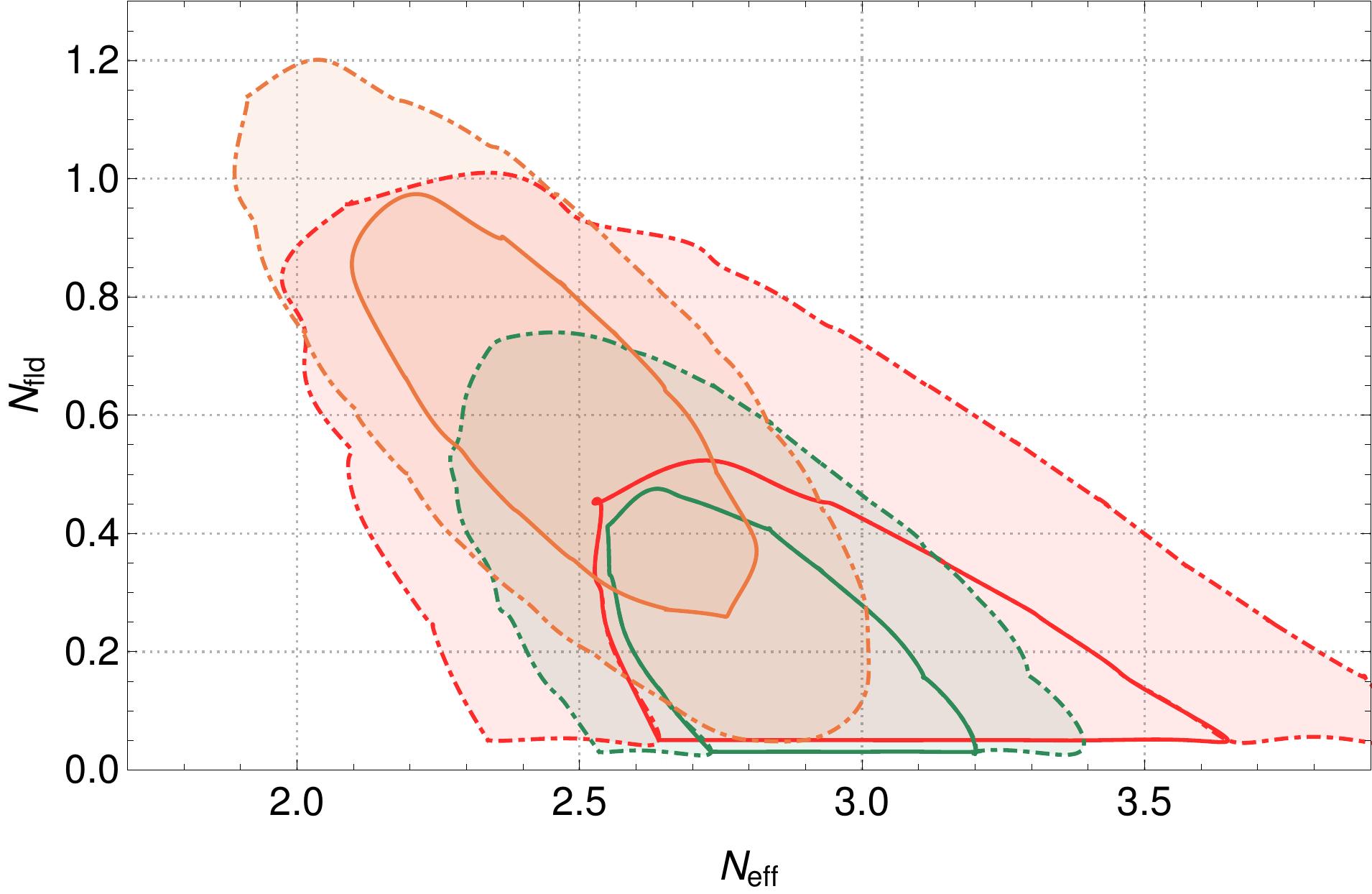}

\vspace{2em}

\includegraphics[width=0.8\textwidth]{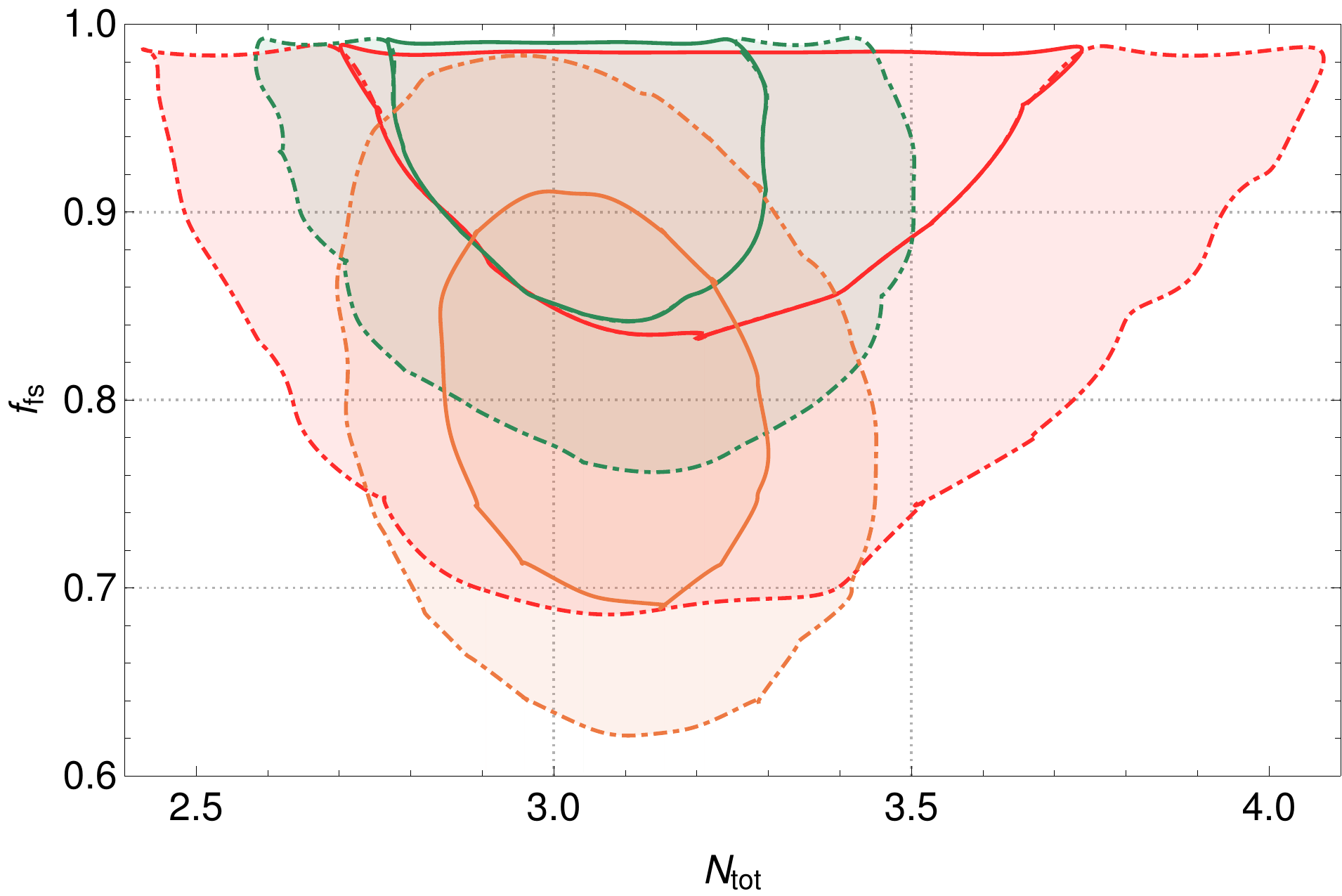}
\caption{\label{fig:2dpresentposteriors}Here we show two different 2d posteriors for three of the five scans ({\color{niceblue} Planck T}, {\color{nicegreen} Planck P+BAO}, and {\color{niceorange} Planck P+BAO+$H_0$+LSS}). The solid lines are 1$\sigma$ contours, and the dot-dashed lines are 2$\sigma$ contours. The posteriors in the top figure exhibit degeneracy between $N_{\rm eff}$ and $N_{\rm fld}$, motivating the parametrization in terms of $N_{\rm tot}$ and $f_{\rm fs}$ in the bottom figure, and demonstrating that the strongest constraints arise on the sum $N_{\rm tot}$. }
\end{figure}

We ran MCMC chains for the following collection of present-day likelihoods: {\color{niceblue} Planck T},  {\color{nicered} Planck P}, {\color{nicegreen} Planck P+BAO}, {\color{mypurple} Planck P+BAO+$H_0$}, and {\color{niceorange} Planck P+BAO+$H_0$+LSS}. We present the results below in table \ref{tab:presentconstraints}, showing either the best-fit value and the 1$\sigma$ posteriors, or the 2$\sigma$ upper bound on parameters. We imposed flat priors on the first block of eight parameters, and derived the remaining five in the second block. We have color-coded the combinations of likelihoods that we used in table \ref{tab:presentconstraints}, and with the same color scheme, we show the 1d posteriors in figure \ref{fig:1dpresentposteriors} and two 2d posteriors in figure \ref{fig:2dpresentposteriors}.

We find a strong degeneracy between $N_{\rm eff}$ and $N_{\rm fld}$, as shown in the upper part of figure \ref{fig:2dpresentposteriors} below. The constraints on the sum $N_{\rm tot}\equiv N_{\rm eff}+N_{\rm fld}$ are considerably more severe than on the difference of these parameters, motivating us to instead parametrize new light species with $N_{\rm tot}$ and $f_{\rm fs}\equiv \frac{N_{\rm eff}}{N_{\rm tot}}$, the fraction of effective neutrinos that are free-streaming. We demonstrate the improved independence of this parametrization in the lower part of figure \ref{fig:2dpresentposteriors} below. Furthermore, specifically for the {\color{nicegreen} Planck P+BAO} run, we produce a full triangle plot in the appendix, in figure \ref{fig:triangle}. The proposed parametrization based on $N_{\rm tot}, f_{\rm fs}$ is also physically motivated: $N_{\rm tot}$ represents the total energy contribution from radiation (except for photons) to the Hubble expansion rate at the time of recombination, while $f_{\rm fs}$ relates to the amplitude of the universal phase shift of the high-$\ell$ peaks in the CMB spectra due to the presence of free-streaming species \cite{Bashinsky:2003tk, Chacko:2015noa}, the damping effects on scalar and tensor modes of CMB, as well as the enhancement effect on matter fluctuation. 
\FloatBarrier

{\renewcommand{\arraystretch}{1.3}
\begin{table}[h]
\small
\centering
\begin{tabular}{|l|c|c|c|c|c|}
 \hline
Param. & {\color{niceblue} Planck T} & {\color{nicered} Planck P} & {\color{nicegreen} Planck P+BAO} & {\color{mypurple} Planck P+BAO}& {\color{niceorange} Planck P+BAO}\\ 
& & & & {\color{mypurple} +$H_0$ }& {\color{niceorange} +$H_0$+LSS}\\ \hline
$100~\Omega_{b }h^2$ & $2.257_{-0.042}^{+0.036}$  &  $2.235_{-0.028}^{+0.025}$ & $2.237_{-0.021}^{+0.021}$  & $2.253_{-0.02}^{+0.019}$ & $2.263_{-0.019}^{+0.019}$ \\
$\Omega_{cdm } h^2$ &$0.1203_{-0.004}^{+0.0039}$ &
$0.1191_{-0.0033}^{+0.0031}$ & 
$0.1192_{-0.0032}^{+0.0031}$& 
$0.1212_{-0.003}^{+0.003}$& 
$0.1176_{-0.0027}^{+0.0026}$\\
$100 \theta_{s }$ &$1.043_{-0.00077}^{+0.00063}$  &
$1.043_{-0.00077}^{+0.00062}$& 
$1.043_{-0.00077}^{+0.00064}$&
$1.043_{-0.00079}^{+0.00065}$&
$1.044_{-0.0008}^{+0.00076}$\\
$\ln 10^{10}A_{s }$  &$3.066_{-0.045}^{+0.036}$   & 
$3.049_{-0.029}^{+0.026}$& 
$3.05_{-0.025}^{+0.024}$ & 
$3.059_{-0.025}^{+0.026}$& 
$3.002_{-0.022}^{+0.018}$\\
$n_{s }$ & $0.9686_{-0.017}^{+0.015}$& 
$0.9601_{-0.0096}^{+0.0094}$ & 
$0.9608_{-0.0081}^{+0.0083}$& 
$0.9659_{-0.0076}^{+0.008}$&
$0.956_{-0.0075}^{+0.0076}$ \\
$\tau$ & $0.06637_{-0.015}^{+0.012}$& 
$0.06699_{-0.012}^{+0.011}$ & 
$0.06649_{-0.012}^{+0.012}$& 
$0.07038_{-0.013}^{+0.012}$& 
 $0.0559_{-0.013}^{+0.0067}$ \\
$N_{\rm eff }$  & $2.916_{-0.36}^{+0.39}$ & 
 $2.795_{-0.23}^{+0.23}$& 
$2.801_{-0.22}^{+0.24}$ & 
$2.9_{-0.22}^{+0.24}$ & 
$2.453_{-0.23}^{+0.23}$  \\
$N_{\rm fld }$ & $<0.7897$ & 
$<0.6181$ &
 $<0.6055$& 
$<0.6852$& 
$0.6138_{-0.24}^{+0.22}$ \\ \hline
$Y_{He}$  & $0.2499_{-0.0043}^{+0.0041}$ & 
$0.2478_{-0.003}^{+0.003}$  &
$0.248_{-0.0026}^{+0.0026}$ & 
$0.2501_{-0.0023}^{+0.0024}$&
$0.2482_{-0.0021}^{+0.0021}$ \\
$H_0$ & $70.1_{-3}^{+2.4}$&
$68.49_{-1.9}^{+1.6}$ &
$68.65_{-1.4}^{+1.3}$ &
$69.82_{-1.2}^{+1.2}$& 
$69.95_{-1.1}^{+1.1}$ \\
$\sigma_8$ & $0.831_{-0.02}^{+0.017}$& 
$0.823_{-0.013}^{+0.012}$ & 
$0.8234_{-0.012}^{+0.012}$& 
$0.8301_{-0.012}^{+0.012}$& 
$0.7971_{-0.0083}^{+0.0077}$\\
$N_{\rm tot}$ & $3.201_{-0.33}^{+0.29}$ & 
$3.046_{-0.22}^{+0.21}$ & 
$3.059_{-0.19}^{+0.19}$ &
$3.211_{-0.18}^{+0.17}$ & 
$3.067_{-0.16}^{+0.15}$ \\
$f_{\rm fs}$ & $>0.7525$ &
$>0.805$ & 
$>0.8069$ & 
$>0.791$ &
$0.8006_{-0.073}^{+0.074}$  \\ \hline 
$\Delta \chi^2$ & $ -0.62$ & $ 0.96$& $ 0.68$ &  $-1.86 $& $-6.84 $ \\ \hline
\end{tabular}
\normalsize
\caption{The results of likelihood analyses for a $\Lambda$CDM$+N_{\rm eff}+N_{\rm fld}$ scan, using various combinations of present-day likelihoods defined above. The first block of parameters had flat priors imposed on them, and the second block were derived. We report a best-fit value and $1\sigma$ posterior almost everywhere, except in a few cases where we instead report a $2\sigma$ upper bound. Finally, we compare $\chi^2$ to a similar scan but using only $\Lambda$CDM, finding no evidence for an improved fit except in the final scan, which has mild evidence for an improved fit. }
\label{tab:presentconstraints}
\end{table}
}

\FloatBarrier

\FloatBarrier

\FloatBarrier

In particular, we note that using only early-time cosmological measurements ({\color{nicegreen} Planck P+BAO}), $N_{\rm tot} < 3.44$ at $2\sigma$, with best fit value of $3.06$ indicating that no more than $0.39$ effective neutrinos may be added (at 2$\sigma$) beyond the SM neutrinos, {\it regardless of whether they are free-streaming or interacting}. In contrast, if we were to consider the $(N_{\rm eff}, N_{\rm fld})$ parametrization, we would obtain $N_{\rm eff}<3.28, N_{\rm fld}<0.62$ at $2\sigma$ which apparently allows more DR in total, resulting from the degeneracy between the two variables as discussed earlier. This data set also imposes constraint that $f_{\rm fs}>0.8$ at 95\% CL, with a preferred value of $\sim 1$.
This conclusion on $N_{\rm tot}$ does not substantially change upon the inclusion of late-time experiments. Adding only a prior on Hubble increases the best-fit value of $N_{\rm tot}$ up to $3.2$, but does not statistically significantly improve the goodness-of-fit over ${\Lambda}$CDM.

Upon further adding LSS priors, though, the best-fit value of $N_{\rm fld}$ shifts away from 0 and up to $0.6$. Equivalently the best-fit value of $f_{\rm fs}$ decreases from 1 to 0.8. However, crucially, {\it the best-fit value of $N_{\rm tot}$ remains at $3.07$}. We further note that this best-fit scenario is preferred at the $2\sigma$ level over $\Lambda$CDM. Apparently the result from the scan suggests the possibility that some fraction of the SM neutrino sector acts as interacting DR. Nevertheless, having $0.6$ neutrinos be interacting does not appear to make physical sense. This can be addressed if a more general parametrization for neutrino interactions is adopted, allowing for weaker interactions beyond the simplified tightly-coupled limit we consider in this work. Earlier attempts in this direction include \cite{Cyr-Racine:2013fsa}. An alternative interpretation could involve DR from a dark sector and non-standard cosmology which dilutes $N_{\rm eff}$ through late entropy release after neutrino decoupling. The apparent larger fraction of $N_{\rm fld}$ given late-time measurements is another curious aspect that may inspire model building involving exotic physics associated with neutrinos and/or a dark sector. The effect of a larger $N_{\rm fld}$ may also be reproduced by a more extended BSM cosmology involving DR and DM interactions, as suggested in \cite{Lesgourgues:2015wza, Chacko:2016kgg}.

In light of the recently recognized tensions between the early- and late-time measurements of $H_0$ and $\sigma_8$, we further explore these late-time priors by explicitly showing the degeneracies between the parameters in the DR sector and the parameters we impose priors on. In particular, it has been recognized that the Hubble expansion rate $H_0$ today
inferred from a fit to the CMB and BAO data based on $\Lambda$CDM paradigm \cite{Ade:2015xua} is smaller than the observed value from direct local 
measurements~\cite{Riess:2011yx, 2013ApJ...766...70S, Riess:2016jrr, Bernal:2016gxb}, 
with a $\sim 3\sigma$ discrepancy; the inferred value of 
$\sigma_8$ is larger by 
$3$--$4\sigma$~\cite{Heymans:2013fya, Ade:2013lmv, MacCrann:2014wfa} 
than the values from late-time measurements such as a weak lensing 
survey~\cite{Fu:2014loa}, CMB lensing~\cite{Ade:2015zua}, and 
Sunyaev-Zeldovich cluster counts~\cite{2012ApJ...755...70R, Hasselfield:2013wf, Ade:2015fva}.

\begin{figure}[h]
\centering
\includegraphics[width=0.49\textwidth]{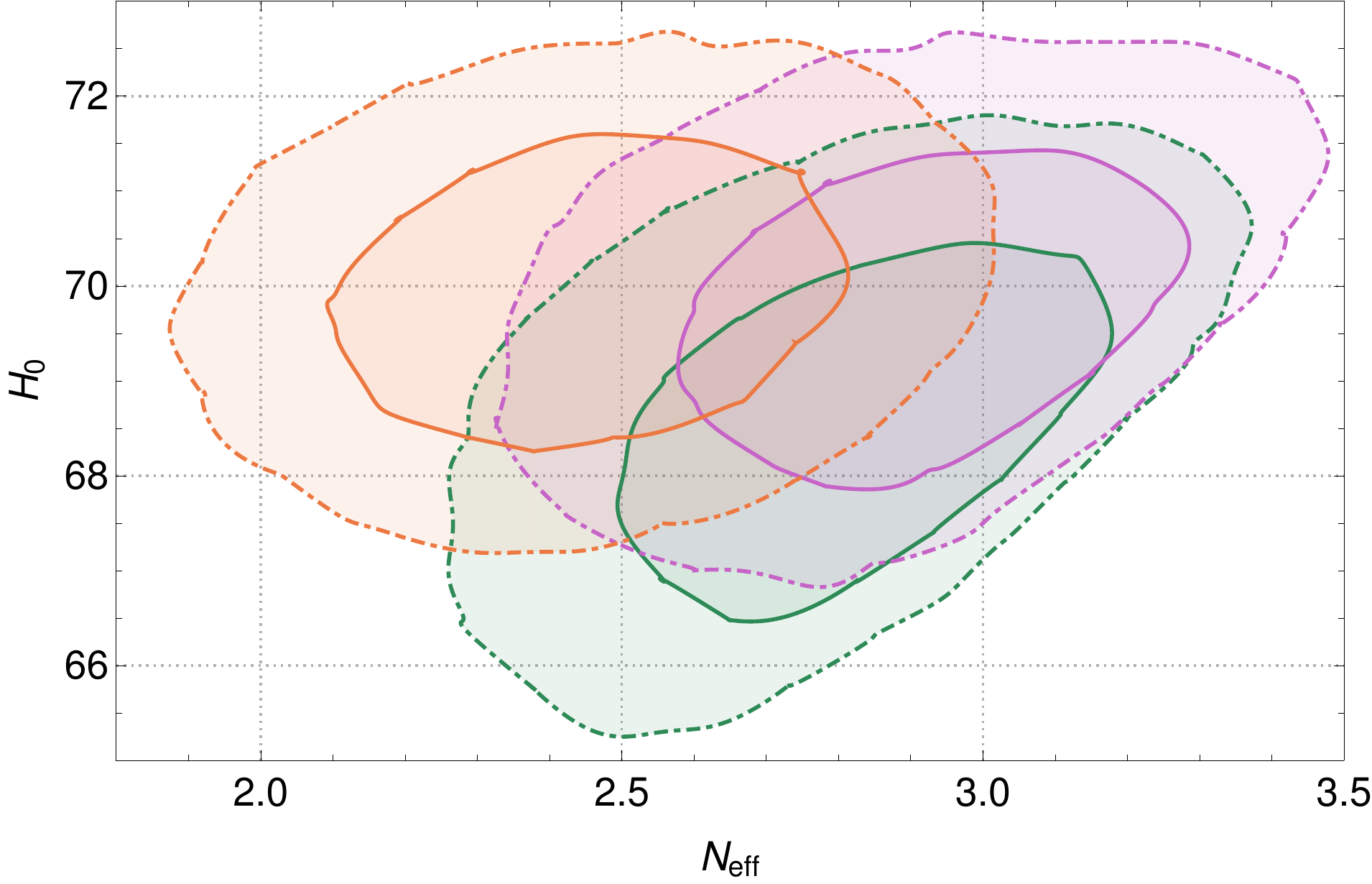}
\includegraphics[width=0.49\textwidth]{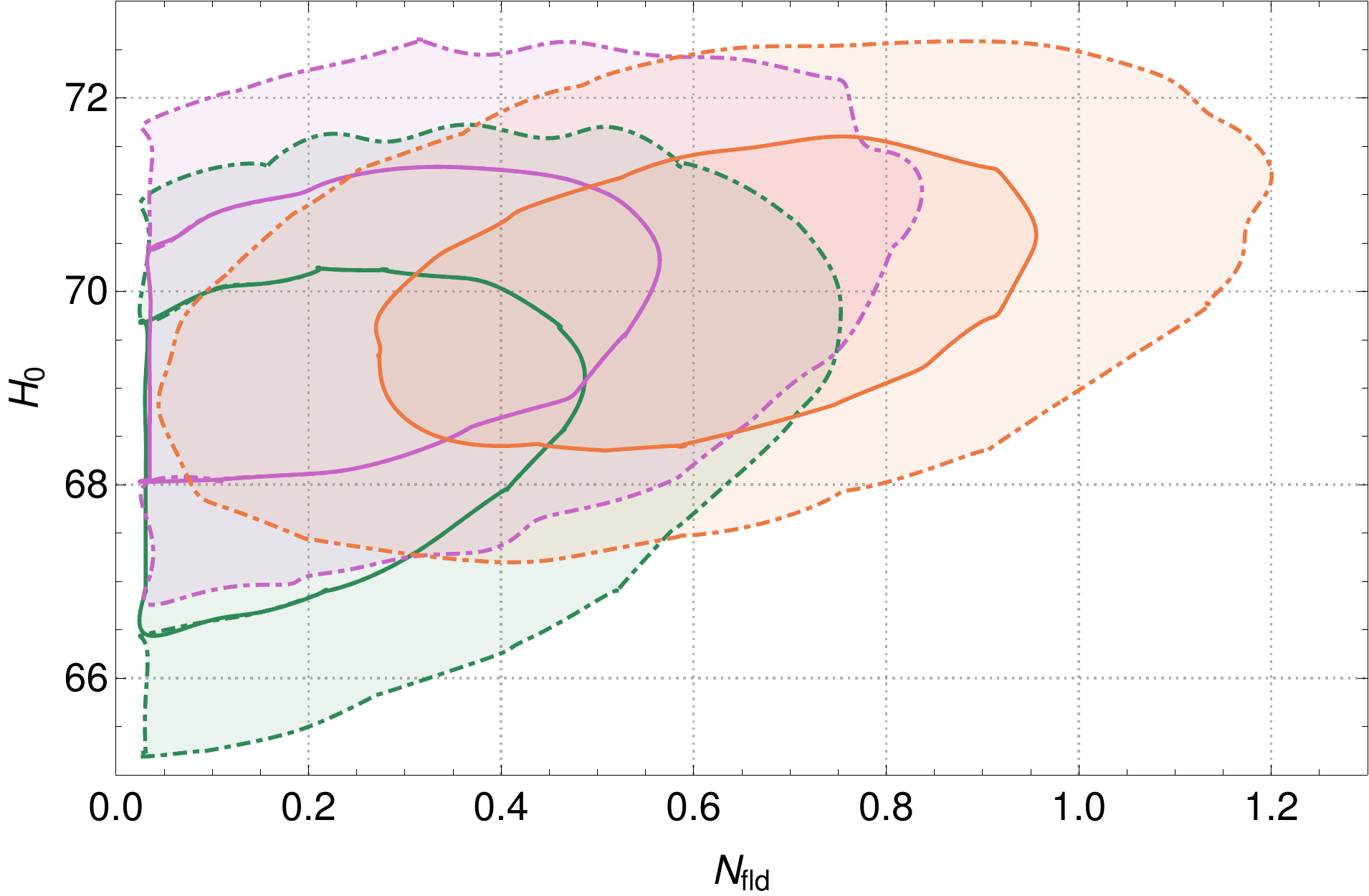}
\includegraphics[width=0.49\textwidth]{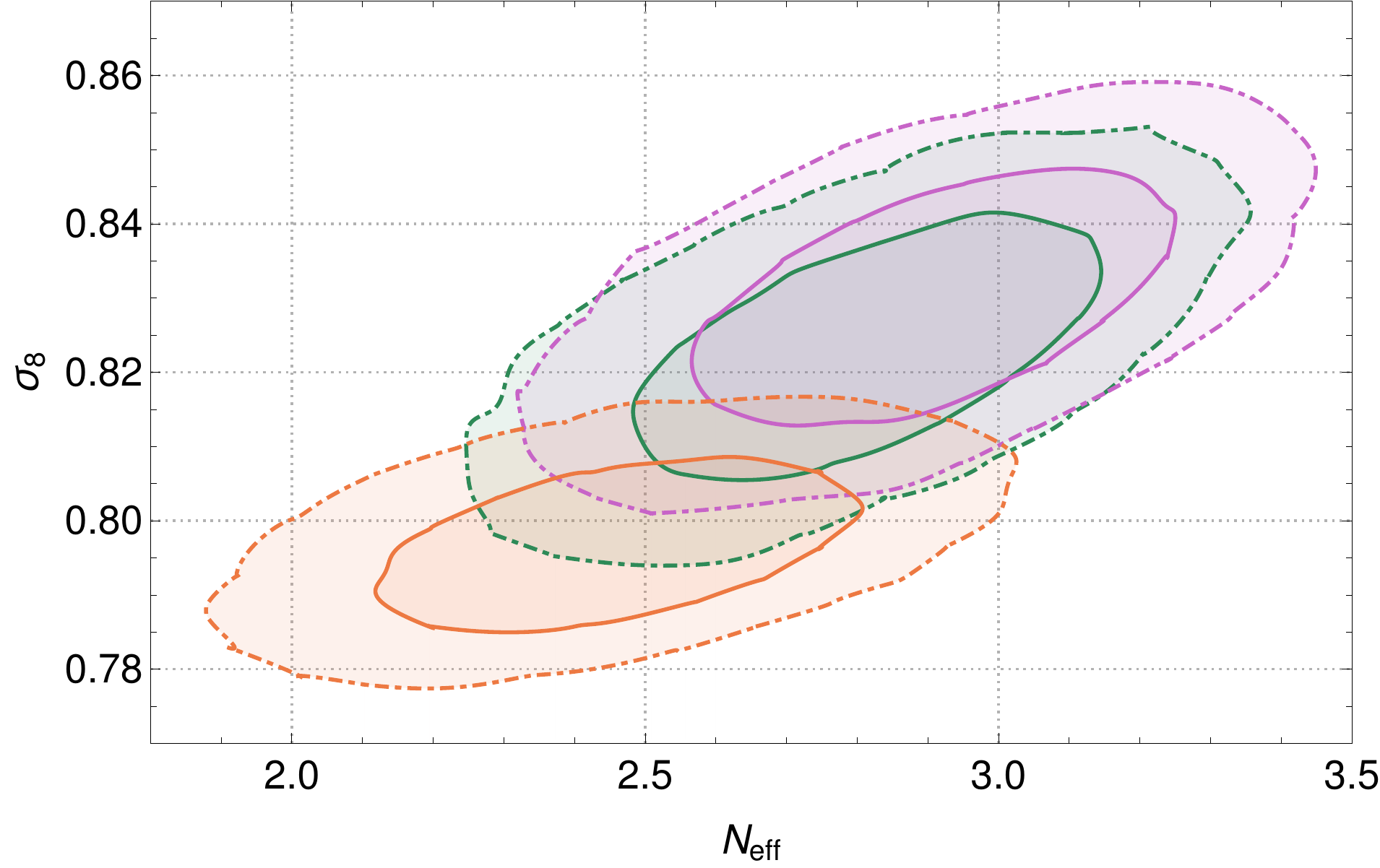}
\includegraphics[width=0.49\textwidth]{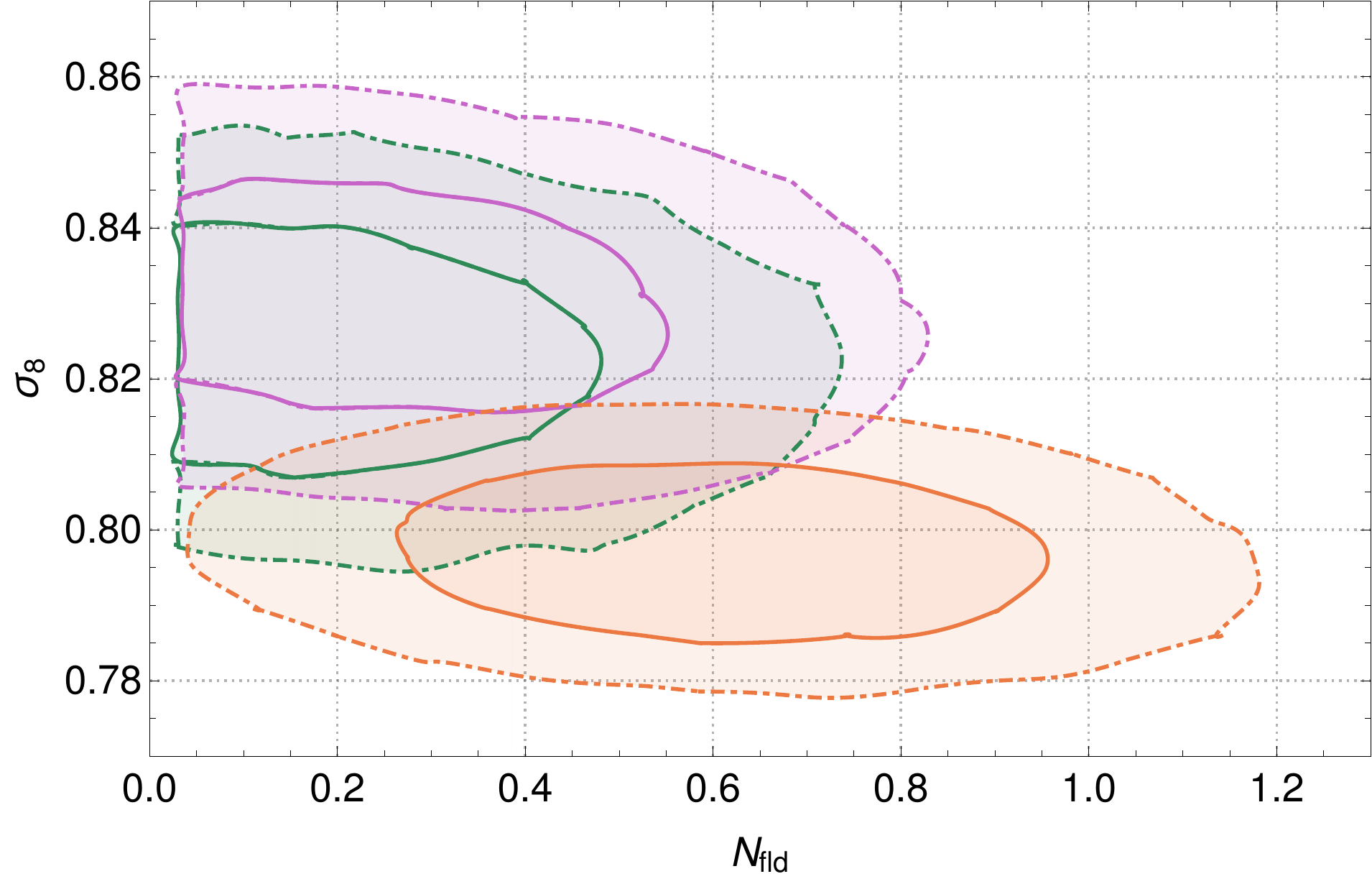}
\caption{\label{fig:2dpresenttensionposteriors}Here we show different 2d posteriors for three of the five scans ({\color{nicegreen} Planck P+BAO}, {\color{mypurple} Planck P+BAO+$H_0$}, and {\color{niceorange} Planck P+BAO+$H_0$+LSS}). The solid lines are 1$\sigma$ contours, and the dot-dashed lines are 2$\sigma$ contours. The posteriors demonstrate how free-streaming DR affects late-time cosmological observables differently than interacting DR does. In the left figures, we explore the effect of $N_{\rm eff}$; in the right figures, $N_{\rm fld}$. In the top figures, we explore the effect on $H_0$; in the bottom figures, $\sigma_8$.} \label{fig:h0s8_now}
\end{figure}

 As shown in figure \ref{fig:h0s8_now}, we find that the inclusion of DR does alleviate the aforementioned tensions in the correct direction (i.e. to increase $H_0$ and decrease $\sigma_8$ compared to fitting CMB with $\Lambda$CDM), although the effect on $\sigma_8$ is much more apparent. Such effect can be understood as follows. As pointed out in earlier literature \cite{Bashinsky:2003tk}, a finite fraction of free-streaming particles in the total radiation density ($R_{\nu}$ as in \cite{Bashinsky:2003tk}, represented by $f_{\rm fs}$ as defined in our paper) causes a super-horizon \textit{perturbation} that affects the gravitational potential and consequently amplifies the matter fluctuation. In contrast, having additional interacting light particles would reduce the fraction of the free-streaming neutrinos in the total radiation energy density, thus suppressing the matter power spectrum. Another factor to consider is that additional relativistic species, either free-streaming or fluid-like, would increase the \textit{background} radiation density, and thus would change fixed CMB variables such as $t_{\rm eq}$ unless matter fluctuation is increased accordingly. For free-streaming neutrinos, the above two factors add up to enhance the amplitude of matter perturbation, while for fluid-like dark radiation, the above factors contribute with opposite signs, thus weakening the net effect on $\sigma_8$ relative to that of free-streaming species. These findings indicate that interacting DR impacts late-time observables substantially differently than free-streaming DR does, suggesting that if a BSM solution to the early vs. late time tension is needed, interacting DR may play a vital role in the resolution of this puzzle. Such a possibility was already suggested in earlier studies involving DM-DR interactions \cite{Buen-Abad:2015ova, Lesgourgues:2015wza, Chacko:2016kgg}.  As shown in figure \ref{fig:h0s8_now}, although the mere inclusion of DR does not fully eliminate the ``anomaly'' related to the $H_0$ and $\sigma_8$ measurements, it does alleviate the tensions in the correct direction, although the effect on $\sigma_8$ is much more apparent. If the early- vs. late-time anomalies become sharper over the coming years, a more extended theory, possibly including DM and DR interactions, may be needed to fully address them, such as in \cite{Lesgourgues:2015wza, Chacko:2016kgg}.

\FloatBarrier

\subsection{$\tau$ Prior}

Here, we study the effects of $\tau$ (the Thomson scattering optical depth) prior on present-day constraints. We again performed a likelihood analysis of the ``Lollipop'' likelihood defined above. Our findings are on the left side of table \ref{tab:lollipopandfuture}. We find no significant evidence for new light species with this prior.

\FloatBarrier

\subsection{Future Constraints}

\begin{figure}[h]
\centering
\includegraphics[width=0.49\textwidth]{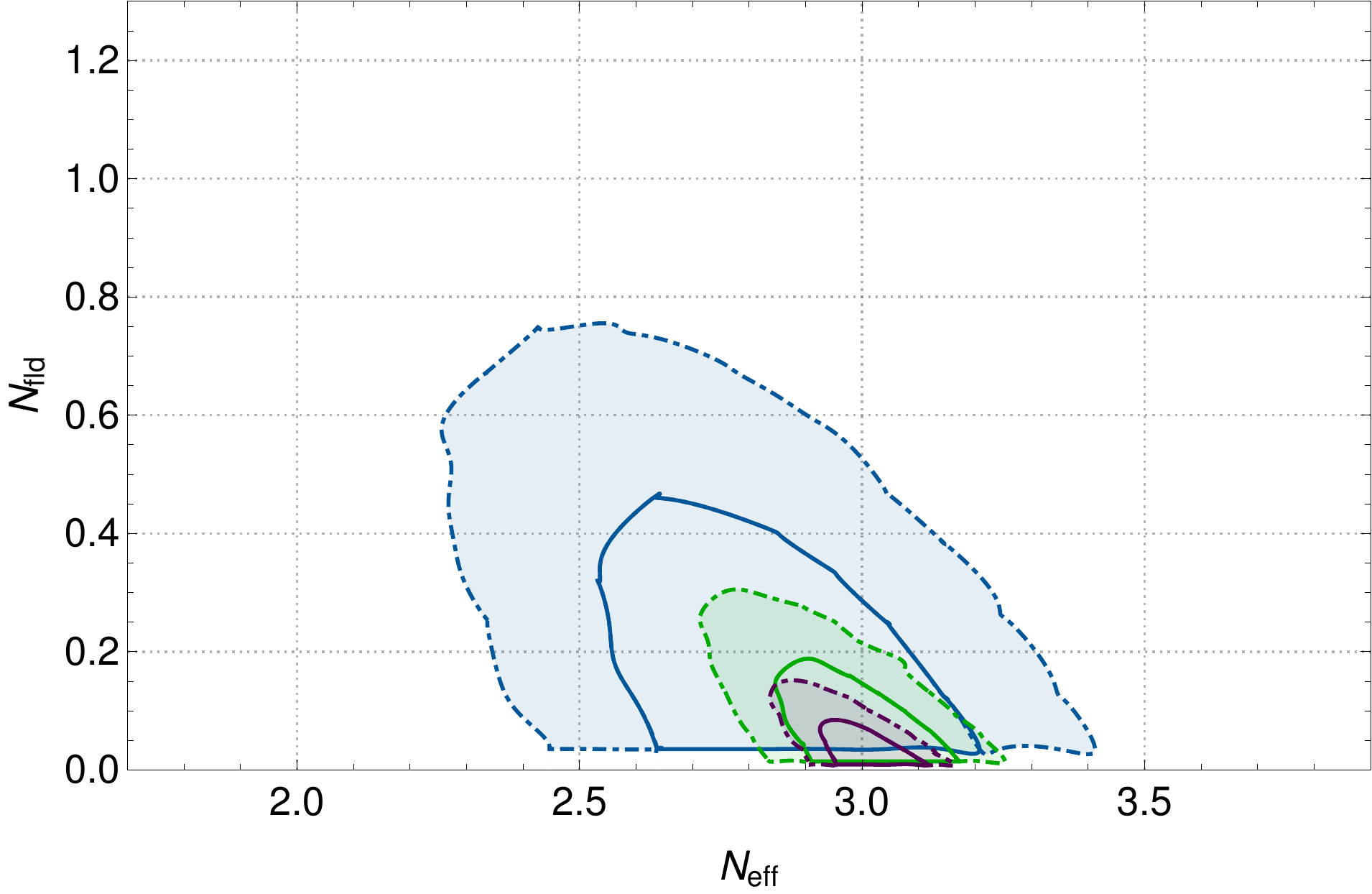}
\includegraphics[width=0.49\textwidth]{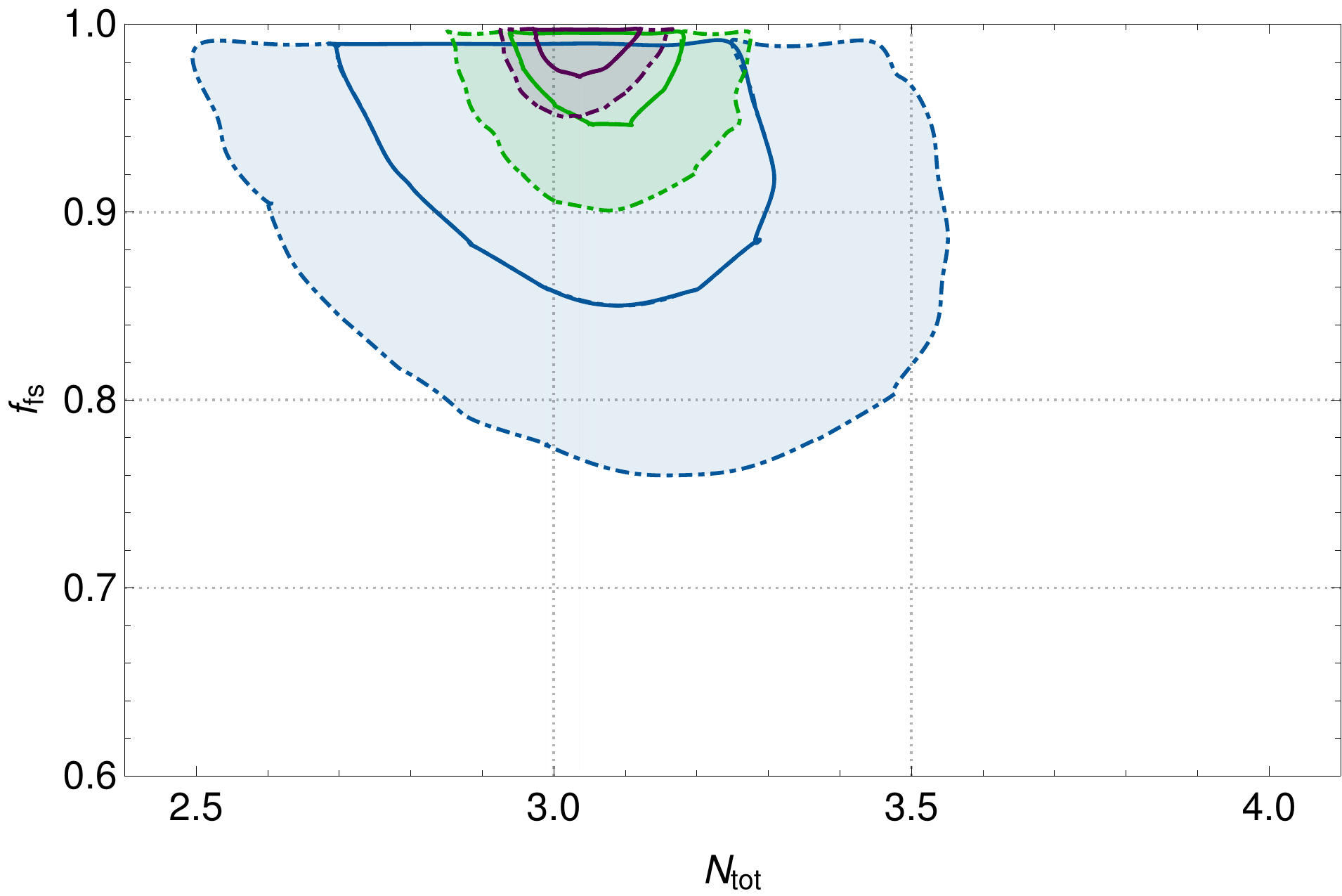}
\caption{\label{fig:futureconstraints}Here we show two different 2d posteriors for two of the three future constraints ({\color{darkergreen} Planck P+S3} and {\color{darkerpurple} Planck P+S4}) alongside one of the present-day constraints ({\color{nicered} Planck P}). The solid lines are 1$\sigma$ contours, and the dot-dashed lines are 2$\sigma$ contours. The left figure shows the conventional parametrization in terms of $N_{\rm eff}$ and $N_{\rm fld}$, whereas the right figure shows our alternative parametrization in terms of $N_{\rm tot}$ and $f_{\rm fs}$.} 
\end{figure}

Finally, we forecast for upcoming experiments. We run scans with the following likelihoods (in chronological order): {\color{darkergreen} Planck P+S3}, { Planck P+Euclid}, and {\color{darkerpurple} Planck P+S4}. Our results are shown on the right side of table \ref{tab:lollipopandfuture}. We also show a comparison of the 2d posteriors for the Stage III and Stage IV results with the present-day constraints (in particular, {\color{nicered} Planck P}) in figure \ref{fig:futureconstraints}.

\FloatBarrier

{\renewcommand{\arraystretch}{1.3}
\begin{table}[h]
\centering
\small
\begin{tabular}{|l|c|}
 \hline
Param & ``Lollipop'' \\ \hline
$100~\Omega_{b }h^2$ & $2.229_{-0.026}^{+0.024}$\\
$\Omega_{cdm }h^2$ & $0.119_{-0.0032}^{+0.0031}$ \\
$100\theta_{s }$ &$1.043_{-0.00073}^{+0.0006}$ \\
$\ln 10^{10}A_{s }$  &$3.039_{-0.023}^{+0.021}$\\
$n_{s }$ & $0.958_{-0.0087}^{+0.0089}$\\
$\tau$ & $0.06077_{-0.011}^{+0.0091}$\\
$N_{\rm eff }$  & $2.783_{-0.21}^{+0.23}$   \\
$N_{\rm fld }$ & $<0.5709$ \\ \hline
$Y_{He}$  &$0.2473_{-0.0028}^{+0.0029}$ \\
$H_0$ & $68.06_{-1.7}^{+1.6}$ \\
$\sigma_8$ & $0.8192_{-0.011}^{+0.011}$  \\
$N_{\rm tot}$ & $3.012_{-0.21}^{+0.2}$ \\
$f_{\rm fs}$ & $>0.8169$ \\ \hline 
\end{tabular} \begin{tabular}{|l|c|c|c|}
 \hline
Param & {\color{darkergreen} Planck P+S3} & { Planck P+Euclid} & {\color{darkerpurple} Planck P+S4} \\ \hline
$100~\Omega_{b }h^2$ & $2.226_{-0.011}^{+0.01}$&  $2.227_{-0.014}^{+0.013}$& $2.225_{-0.0051}^{+0.0051}$\\
$\Omega_{cdm }h^2$ &$0.1201_{-0.0014}^{+0.0014}$  & $0.1226_{-0.0022}^{+0.002}$&$0.1196_{-0.0011}^{+0.001}$\\
$100\theta_{s }$ & $1.041_{-0.00031}^{+0.00024}$ & $1.042_{-0.00047}^{+0.00044}$& $1.041_{-0.00017}^{+0.00014}$\\
$\ln 10^{10}A_{s }$  & $3.075_{-0.016}^{+0.017}$ & $3.079_{-0.012}^{+0.011}$& $3.089_{-0.012}^{+0.012}$\\
$n_{s }$ & $0.964_{-0.0045}^{+0.0047}$ &$0.9637_{-0.0037}^{+0.0035}$ & $0.9644_{-0.0037}^{+0.0037}$\\
$\tau$ &  $0.07204_{-0.0085}^{+0.0093}$ & $0.07338_{-0.0077}^{+0.0074}$& $0.07813_{-0.0068}^{+0.0068}$\\
$N_{\rm eff }$  & $2.973_{-0.094}^{+0.11}$  & $3.024_{-0.11}^{+0.12}$ &   $2.995_{-0.056}^{+0.071}$\\
$N_{\rm fld }$ &  $<0.2461$&$ <0.355$ &  $<0.1226$\\ \hline
$Y_{He}$  & $0.2481_{-0.0012}^{+0.0011}$ & $0.2493_{-0.0015}^{+0.0014}$&  $0.2477_{-0.00067}^{+0.00064}$\\
$H_0$ & $67.79_{-0.73}^{+0.71}$ & $68.27_{-0.64}^{+0.54}$& $67.56_{-0.46}^{+0.46}$\\
$\sigma_8$ & $0.8359_{-0.007}^{+0.0071}$ & $0.8441_{-0.0037}^{+0.0026}$& $0.8408_{-0.0046}^{+0.0045}$\\ 
$N_{\rm tot}$ & $3.067_{-0.086}^{+0.082}$ & $3.157_{-0.11}^{+0.1}$& $3.04_{-0.048}^{+0.047}$ \\
$f_{\rm fs}$ & $>0.92$  & $>0.891$& $>0.9595$ \\ \hline 
\end{tabular}
\normalsize
\caption{Left: The results of a likelihood analysis for a $\Lambda$CDM$+N_{\rm eff}+N_{\rm fld}$ scan using the ``Lollipop'' combination of likelihoods defined above, including a relatively low prior on $\tau$. Right: The forecasted results of likelihood analyses for a $\Lambda$CDM$+N_{\rm eff}+N_{\rm fld}$ scan, using the various mock likelihoods defined above, in chronological order. Both: the first block of parameters had flat priors imposed on them, and the second block were derived. We report a best-fit value and $1\sigma$ posterior almost everywhere, except for $N_{\rm fld}$ and $f_{\rm fs}$ where we instead report a $2\sigma$ bound.}
\label{tab:lollipopandfuture}
\end{table}
}
\FloatBarrier

\begin{figure}[h]
\centering
\includegraphics[width=0.49\textwidth]{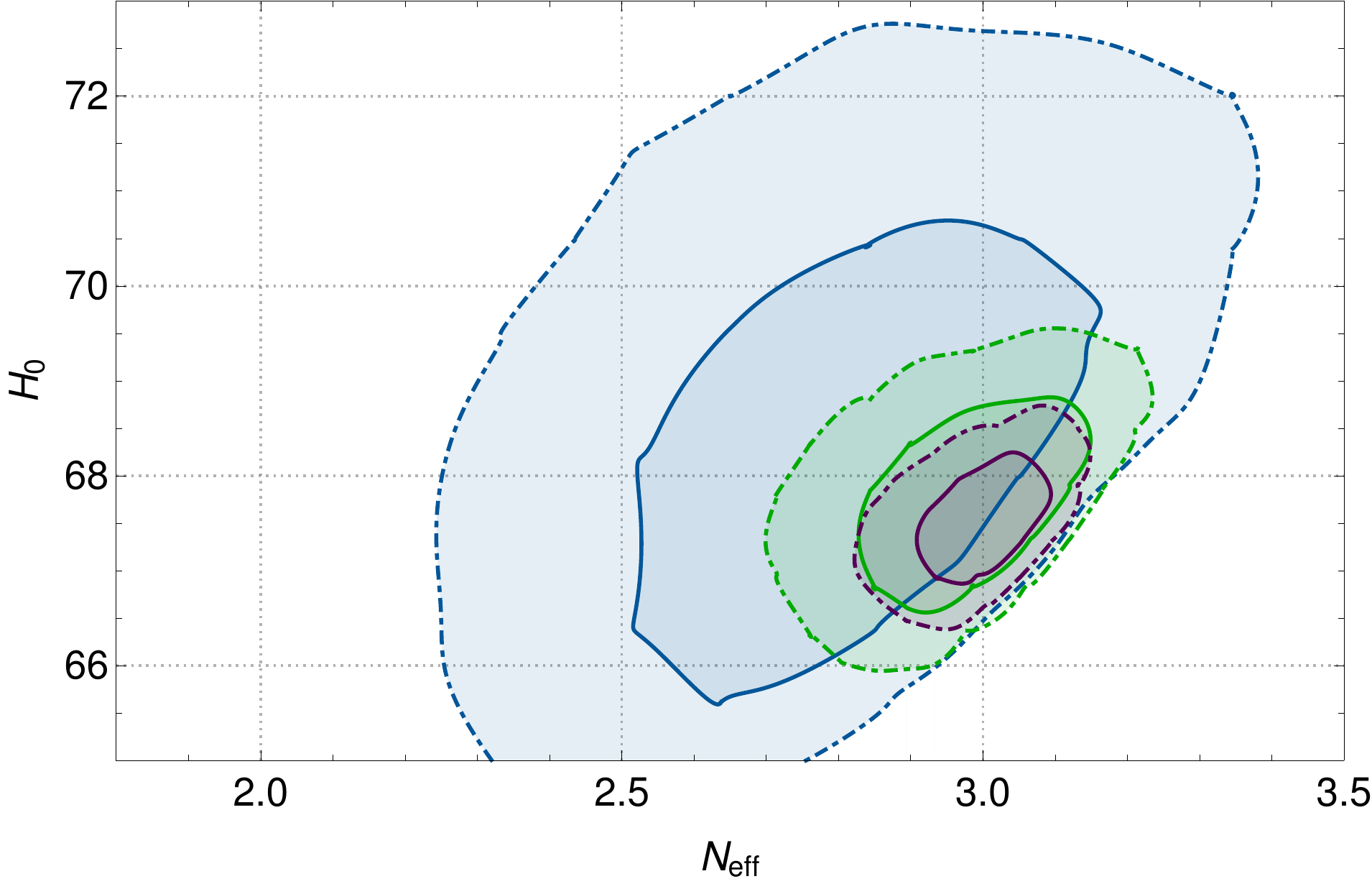}
\includegraphics[width=0.49\textwidth]{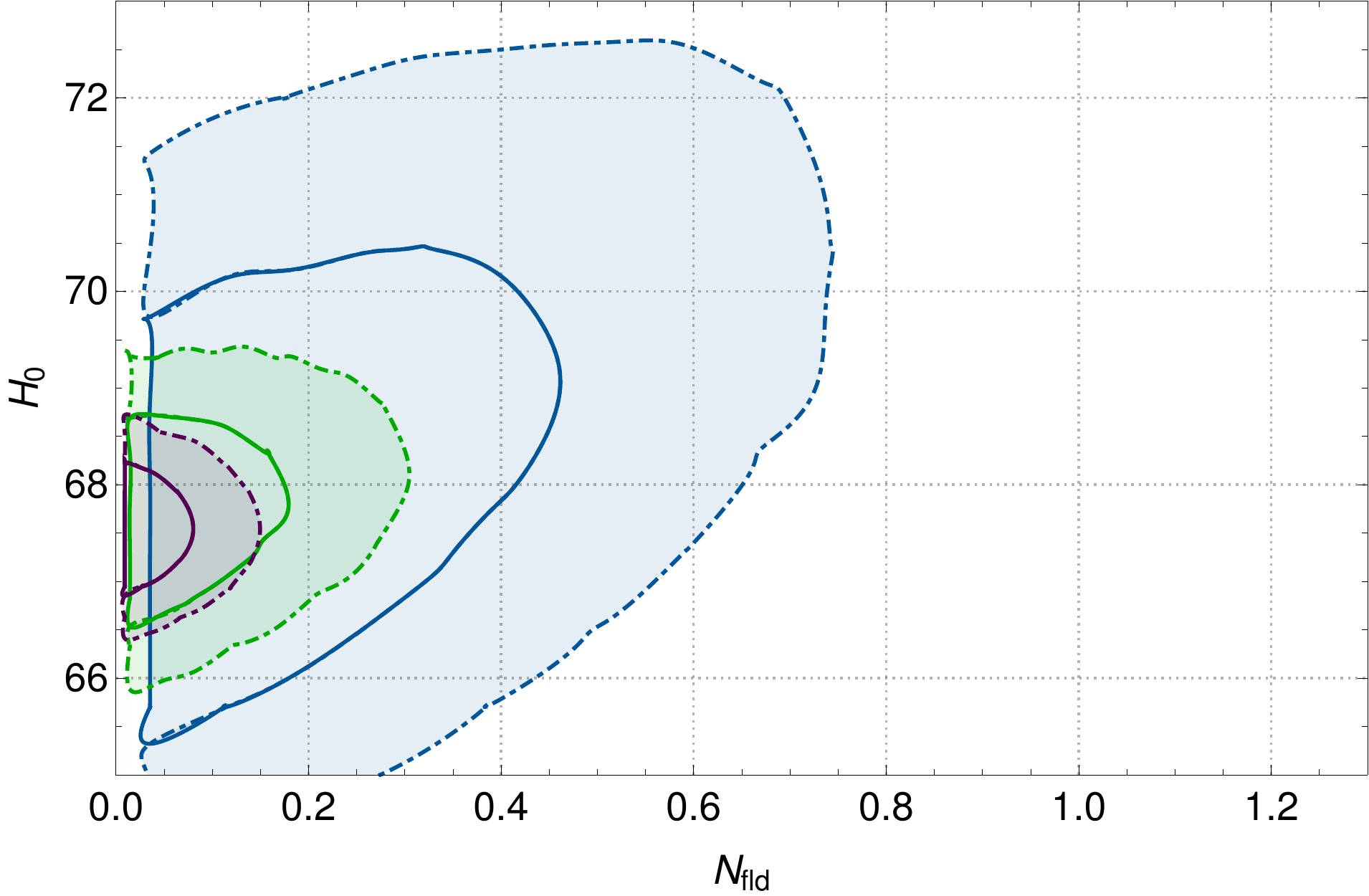}
\includegraphics[width=0.49\textwidth]{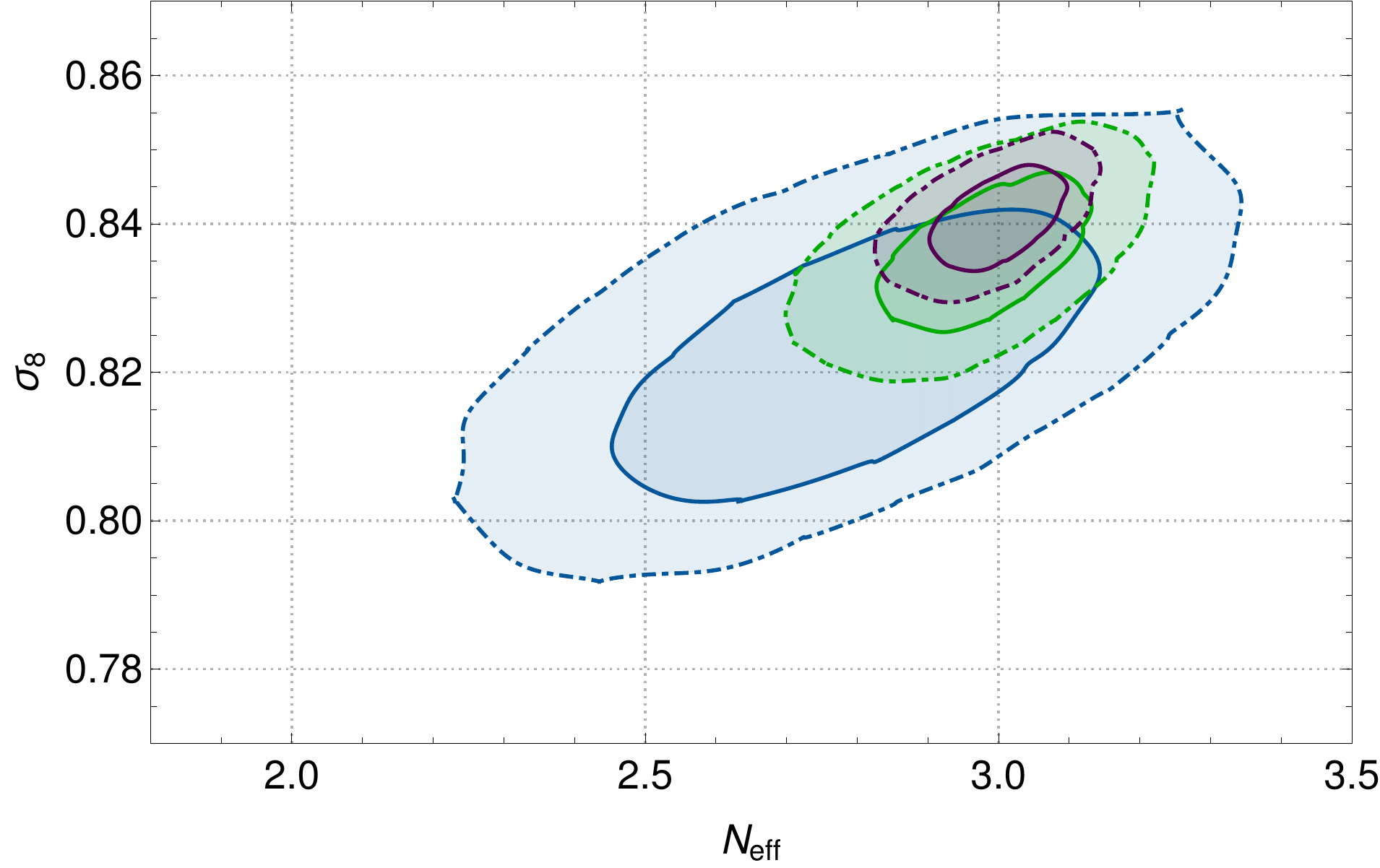}
\includegraphics[width=0.49\textwidth]{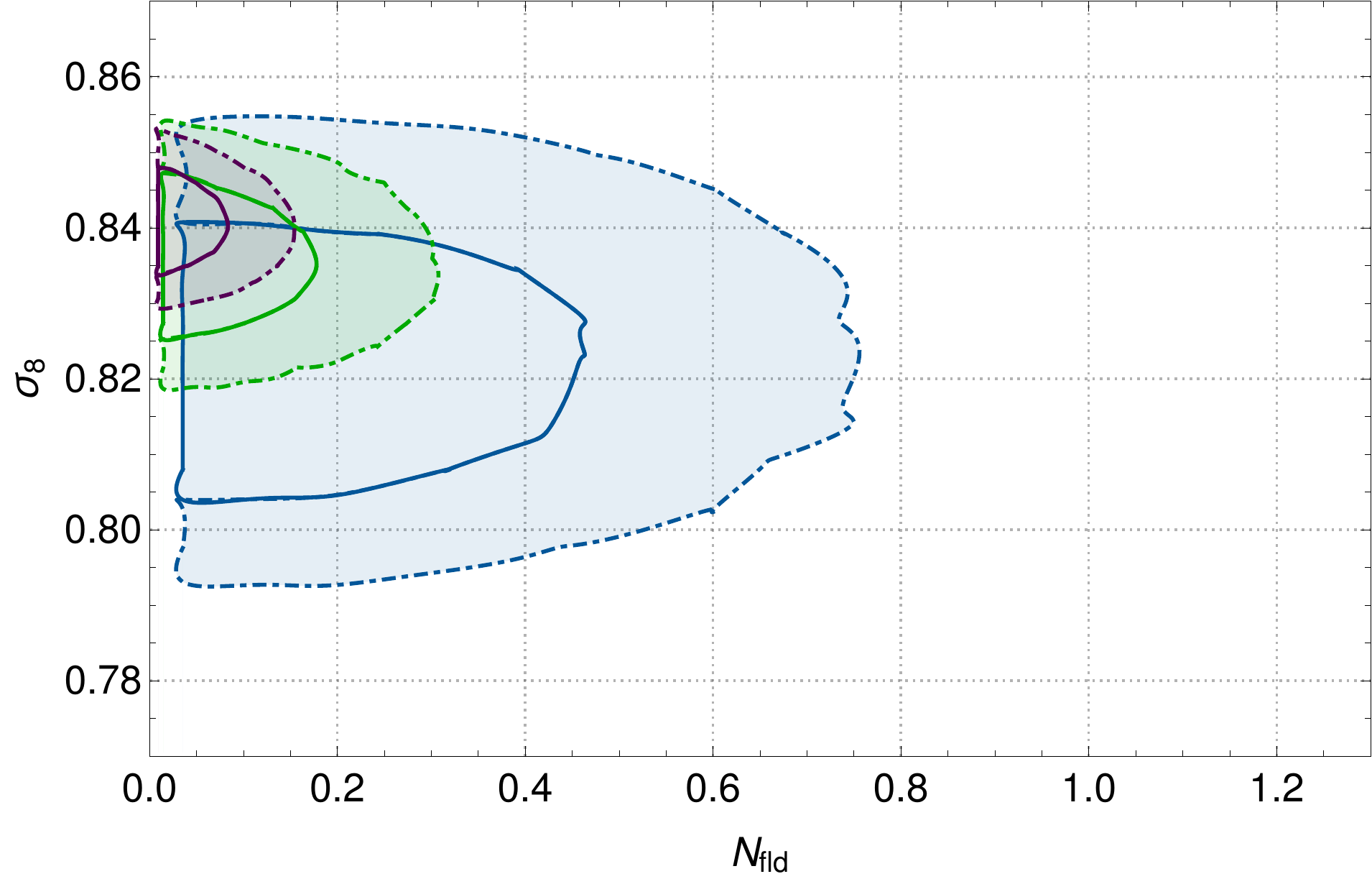}
\caption{\label{fig:2dfuturetensionposteriors}Here we show two different 2d posteriors for one present-day and two mock scans ({\color{nicered} Planck P}, {\color{darkergreen} Planck P+S3}, and {\color{darkerpurple} Planck P+S4}). The solid lines are 1$\sigma$ contours, and the dot-dashed lines are 2$\sigma$ contours. The posteriors demonstrate how free-streaming DR will affect late-time cosmological observables differently than interacting DR will. In the left figures, we explore the effect of $N_{\rm eff}$; in the right figures, $N_{\rm fld}$. In the top figures, we explore the effect on $H_0$; in the bottom figures, $\sigma_8$.} \label{fig: h0s8_forecast1}
\end{figure}

\begin{figure}[h]
\centering
\includegraphics[width=0.49\textwidth]{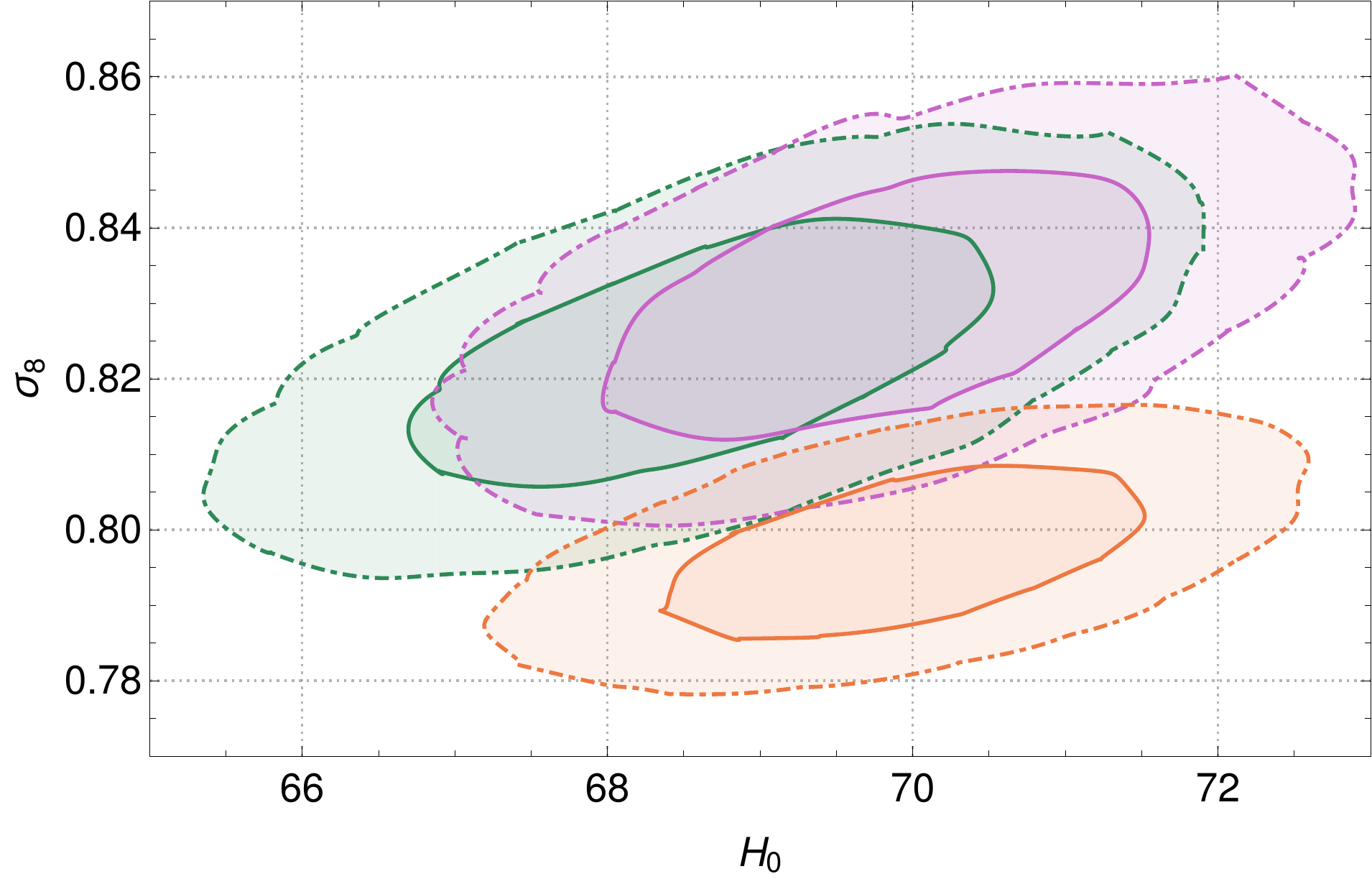}
\includegraphics[width=0.49\textwidth]{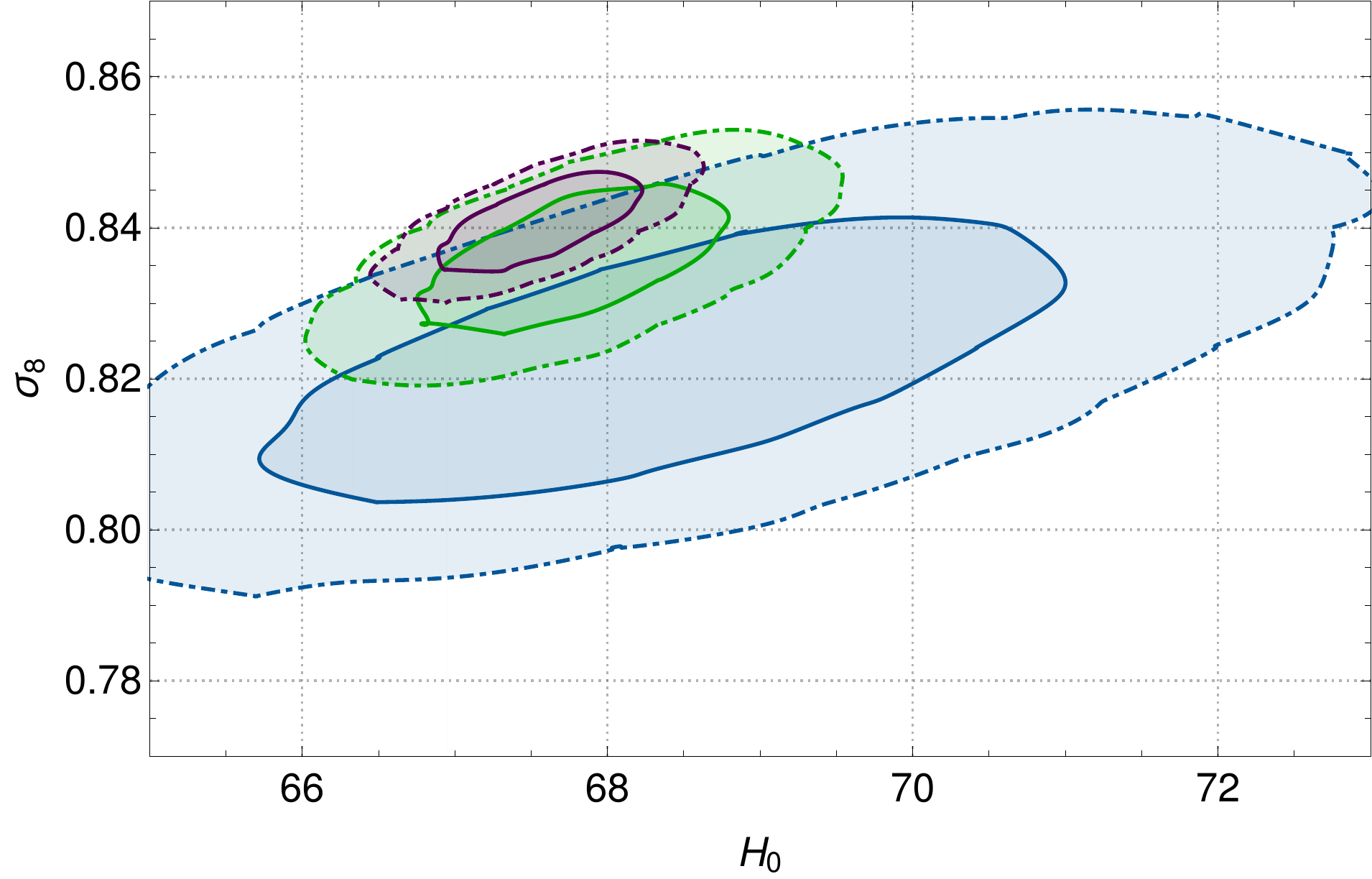}
\caption{\label{fig:2dsig8H0posteriors}Here we show $H_0$-$\sigma_8$ 2d posteriors for six scans. On the left, we show present results with {\color{nicegreen} Planck P+BAO}, {\color{mypurple} Planck P+BAO+$H_0$}, and {\color{niceorange} Planck P+BAO+$H_0$+LSS}. On the right, we show the present {\color{nicered} Planck P} contrasted with the forecasted {\color{darkergreen} Planck P+S3} and {\color{darkerpurple} Planck P+S4}. The solid lines are 1$\sigma$ contours, and the dot-dashed lines are 2$\sigma$ contours.} \label{fig: h0s8_forecast2}
\end{figure}

\FloatBarrier
In figure \ref{fig: h0s8_forecast1}, \ref{fig: h0s8_forecast2}, we also demonstrate future correlations between $H_0$/$\sigma_8$ and DR parameters. Again we can see that interacting DR affects the fitting of $\sigma_8$ in a very different way from the free-streaming species. We also see that future experiments will be able to significantly improve the precision of $\sigma_8-H_0$ fitting.

\FloatBarrier

\FloatBarrier

\section{Conclusions}
\label{sec:conclusions}

The presence of light particles at the recombination epoch is well-motivated by a variety of BSM particle physics considerations. In this paper, we studied both present-day and future constraints on a cosmology with both free-streaming and self-interacting new light species, finding present-day constraints from early-time data ({\color{nicegreen} Planck P+BAO}) of $N_{\rm eff} \sim 2.8\pm 0.2$ at 1$\sigma$ and $N_{\rm fld}< 0.6$ at 2$\sigma$. We motivated an alternate parameterization of the new light degrees of freedom in terms of $N_{\rm tot}$ and $f_{\rm fs}$ which are less degenerate, with more independent, transparent physical meanings. In these new variables, we find a present-day constraint from early-time data of $N_{\rm tot} = 3.06\pm 0.19$ at 1$\sigma$, suggesting that any DR (either interacting or free-streaming) beyond the SM neutrinos has to satisfy $\Delta N_{\rm tot}<0.39$ at 2$\sigma$. We explored the tension with late-time experiments (LSS, direct $H_0$ measurements), finding 2$\sigma$ evidence for having a fraction of SM neutrinos be self-interacting, alleviating the tension between early-time measurements of the CMB+BAO and late-time measurements of large-scale structure and Hubble expansion rate. Finally, we explored how these sensitivities will improve over the next decade as more cosmological data becomes available.

Interesting future directions to explore include: a more general parametrization of the interactions of SM neutrinos and BSM dark radiations beyond the tightly-coupled limit as considered in this work, and how the observational data constrains such possibilities. If the slight preference of having interacting neutrinos or DR as suggested from our analysis gets strengthened with upcoming data, this could inspire further studies on non-standard neutrino physics, non-standard cosmology involving DR possibly interacting with DM, both in terms of model-building and in terms of complementary phenomenology such as implications for astrophysical probes for DM interactions as well as neutrino experiments on ground.

{\bf Acknowledgements:} We are grateful to Daniel Baumann, Dan Green, Matt Johnson, Joel Meyers, and Kendrick Smith for helpful discussions and comments. We are also grateful to Masha Baryakhtar for collaboration during the early stages of this project, and to the Perimeter Institute for time on the HPC cluster. Research at Perimeter Institute is supported by the Government of Canada through Industry Canada and by the Province of Ontario through the Ministry of Economic Development and Innovation. The research of KS is supported in part by a Natural Sciences and Engineering Research Council of Canada (NSERC) Discovery Grant.  KS gratefully acknowledges the hospitality of the Institute for Advanced Study, where this research was initiated, and support from the Friends of the Institute for Advanced Study while in residence there.
\FloatBarrier

\newpage

\appendix

\section{Additional Posteriors}

\FloatBarrier

\begin{figure}[h]
\centering
\includegraphics[width=\textwidth]{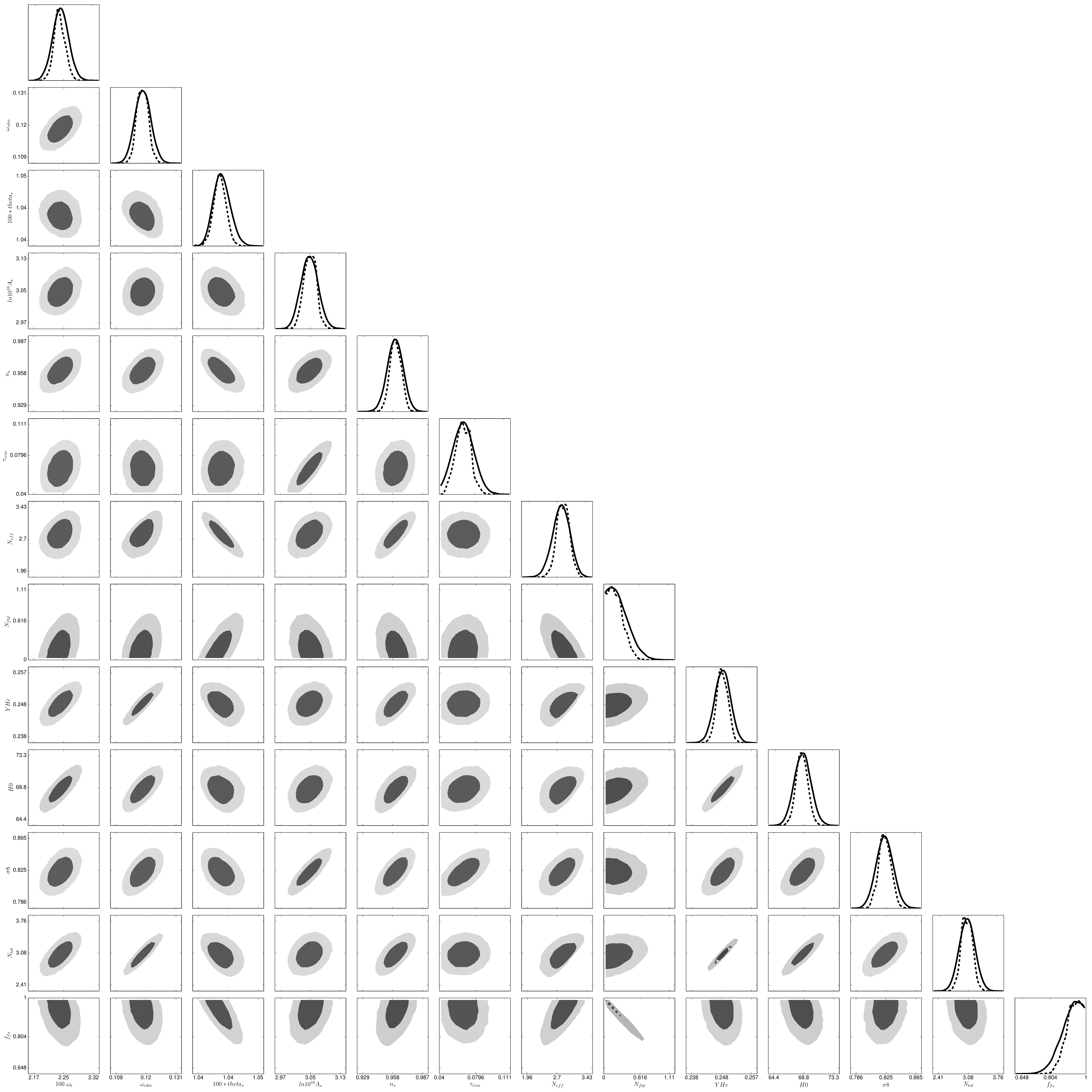}
\caption{\label{fig:triangle}Here we show the 2d posteriors in the {\color{nicegreen} Planck P+BAO} run. The dark gray regions are the 1$\sigma$ allowed regions, and the light gray regions are the 2$\sigma$ allowed regions. These plots illustrate correlations between the various cosmological parameters upon digital zoom-in.}
\end{figure}

\bibliography{ref_Nfld}{}

\providecommand{\href}[2]{#2}\begingroup\raggedright\begin{thebibliography}{10}

\bibitem{Jungman:1995bz}
G.~Jungman, M.~Kamionkowski, A.~Kosowsky and D.~N. Spergel, \emph{{Cosmological
  parameter determination with microwave background maps}},
  \href{http://dx.doi.org/10.1103/PhysRevD.54.1332}{\emph{Phys. Rev.} {\bf D54}
  (1996) 1332--1344}, [\href{http://arxiv.org/abs/astro-ph/9512139}{{\tt
  astro-ph/9512139}}].

\bibitem{Dolgov:2002wy}
A.~D. Dolgov, \emph{{Neutrinos in cosmology}},
  \href{http://dx.doi.org/10.1016/S0370-1573(02)00139-4}{\emph{Phys. Rept.}
  {\bf 370} (2002) 333--535}, [\href{http://arxiv.org/abs/hep-ph/0202122}{{\tt
  hep-ph/0202122}}].

\bibitem{Steigman:2001px}
G.~Steigman, \emph{{Precision neutrino counting}},
  \href{http://arxiv.org/abs/astro-ph/0108148}{{\tt astro-ph/0108148}}.

\bibitem{Lopez:1998aq}
R.~E. Lopez, S.~Dodelson, A.~Heckler and M.~S. Turner, \emph{{Precision
  detection of the cosmic neutrino background}},
  \href{http://dx.doi.org/10.1103/PhysRevLett.82.3952}{\emph{Phys. Rev. Lett.}
  {\bf 82} (1999) 3952--3955},
  [\href{http://arxiv.org/abs/astro-ph/9803095}{{\tt astro-ph/9803095}}].

\bibitem{Nakayama:2010vs}
K.~Nakayama, F.~Takahashi and T.~T. Yanagida, \emph{{A theory of extra
  radiation in the Universe}},
  \href{http://dx.doi.org/10.1016/j.physletb.2011.02.013}{\emph{Phys. Lett.}
  {\bf B697} (2011) 275--279}, [\href{http://arxiv.org/abs/1010.5693}{{\tt
  1010.5693}}].

\bibitem{Fischler:2010xz}
W.~Fischler and J.~Meyers, \emph{{Dark Radiation Emerging After Big Bang
  Nucleosynthesis?}},
  \href{http://dx.doi.org/10.1103/PhysRevD.83.063520}{\emph{Phys. Rev.} {\bf
  D83} (2011) 063520}, [\href{http://arxiv.org/abs/1011.3501}{{\tt
  1011.3501}}].

\bibitem{Chacko:2015noa}
Z.~Chacko, Y.~Cui, S.~Hong and T.~Okui, \emph{{Hidden dark matter sector, dark
  radiation, and the CMB}},
  \href{http://dx.doi.org/10.1103/PhysRevD.92.055033}{\emph{Phys. Rev.} {\bf
  D92} (2015) 055033}, [\href{http://arxiv.org/abs/1505.04192}{{\tt
  1505.04192}}].

\bibitem{Ade:2015xua}
{\scshape Planck} collaboration, P.~A.~R. Ade et~al., \emph{{Planck 2015
  results. XIII. Cosmological parameters}},
  \href{http://arxiv.org/abs/1502.01589}{{\tt 1502.01589}}.

\bibitem{Brust:2013xpv}
C.~Brust, D.~E. Kaplan and M.~T. Walters, \emph{{New Light Species and the
  CMB}}, \href{http://dx.doi.org/10.1007/JHEP12(2013)058}{\emph{JHEP} {\bf 12}
  (2013) 058}, [\href{http://arxiv.org/abs/1303.5379}{{\tt 1303.5379}}].

\bibitem{Stebor:2016hgt}
N.~Stebor et~al., \emph{{The Simons Array CMB polarization experiment}},
  \href{http://dx.doi.org/10.1117/12.2233103}{\emph{Proc. SPIE Int. Soc. Opt.
  Eng.} {\bf 9914} (2016) 99141H}.

\bibitem{Abazajian:2016yjj}
{\scshape CMB-S4} collaboration, K.~N. Abazajian et~al., \emph{{CMB-S4 Science
  Book, First Edition}},  \href{http://arxiv.org/abs/1610.02743}{{\tt
  1610.02743}}.

\bibitem{Wu:2014hta}
W.~L.~K. Wu, J.~Errard, C.~Dvorkin, C.~L. Kuo, A.~T. Lee, P.~McDonald et~al.,
  \emph{{A Guide to Designing Future Ground-based Cosmic Microwave Background
  Experiments}},
  \href{http://dx.doi.org/10.1088/0004-637X/788/2/138}{\emph{Astrophys. J.}
  {\bf 788} (2014) 138}, [\href{http://arxiv.org/abs/1402.4108}{{\tt
  1402.4108}}].

\bibitem{Errard:2015cxa}
J.~Errard, S.~M. Feeney, H.~V. Peiris and A.~H. Jaffe, \emph{{Robust forecasts
  on fundamental physics from the foreground-obscured, gravitationally-lensed
  CMB polarization}},
  \href{http://dx.doi.org/10.1088/1475-7516/2016/03/052}{\emph{JCAP} {\bf 1603}
  (2016) 052}, [\href{http://arxiv.org/abs/1509.06770}{{\tt 1509.06770}}].

\bibitem{Adshead:2016xxj}
P.~Adshead, Y.~Cui and J.~Shelton, \emph{{Chilly Dark Sectors and Asymmetric
  Reheating}}, \href{http://dx.doi.org/10.1007/JHEP06(2016)016}{\emph{JHEP}
  {\bf 06} (2016) 016}, [\href{http://arxiv.org/abs/1604.02458}{{\tt
  1604.02458}}].

\bibitem{Chacko:2003dt}
Z.~Chacko, L.~J. Hall, T.~Okui and S.~J. Oliver, \emph{{CMB signals of neutrino
  mass generation}},
  \href{http://dx.doi.org/10.1103/PhysRevD.70.085008}{\emph{Phys. Rev.} {\bf
  D70} (2004) 085008}, [\href{http://arxiv.org/abs/hep-ph/0312267}{{\tt
  hep-ph/0312267}}].

\bibitem{Friedland:2007vv}
A.~Friedland, K.~M. Zurek and S.~Bashinsky, \emph{{Constraining Models of
  Neutrino Mass and Neutrino Interactions with the Planck Satellite}},
  \href{http://arxiv.org/abs/0704.3271}{{\tt 0704.3271}}.

\bibitem{Cyr-Racine:2013jua}
F.-Y. Cyr-Racine and K.~Sigurdson, \emph{{Limits on Neutrino-Neutrino
  Scattering in the Early Universe}},
  \href{http://dx.doi.org/10.1103/PhysRevD.90.123533}{\emph{Phys. Rev.} {\bf
  D90} (2014) 123533}, [\href{http://arxiv.org/abs/1306.1536}{{\tt
  1306.1536}}].

\bibitem{Kolb:1985bf}
E.~W. Kolb, D.~Seckel and M.~S. Turner, \emph{{The Shadow World}},
  \href{http://dx.doi.org/10.1038/314415a0}{\emph{Nature} {\bf 314} (1985)
  415--419}.

\bibitem{Hodges:1993yb}
H.~M. Hodges, \emph{{Mirror baryons as the dark matter}},
  \href{http://dx.doi.org/10.1103/PhysRevD.47.456}{\emph{Phys. Rev.} {\bf D47}
  (1993) 456--459}.

\bibitem{Feng:2008mu}
J.~L. Feng, H.~Tu and H.-B. Yu, \emph{{Thermal Relics in Hidden Sectors}},
  \href{http://dx.doi.org/10.1088/1475-7516/2008/10/043}{\emph{JCAP} {\bf 0810}
  (2008) 043}, [\href{http://arxiv.org/abs/0808.2318}{{\tt 0808.2318}}].

\bibitem{Ackerman:mha}
L.~Ackerman, M.~R. Buckley, S.~M. Carroll and M.~Kamionkowski, \emph{{Dark
  Matter and Dark Radiation}},
  \href{http://dx.doi.org/10.1103/PhysRevD.79.023519,
  10.1142/9789814293792_0021}{\emph{Phys. Rev.} {\bf D79} (2009) 023519},
  [\href{http://arxiv.org/abs/0810.5126}{{\tt 0810.5126}}].

\bibitem{Kaplan:2009de}
D.~E. Kaplan, G.~Z. Krnjaic, K.~R. Rehermann and C.~M. Wells, \emph{{Atomic
  Dark Matter}},
  \href{http://dx.doi.org/10.1088/1475-7516/2010/05/021}{\emph{JCAP} {\bf 1005}
  (2010) 021}, [\href{http://arxiv.org/abs/0909.0753}{{\tt 0909.0753}}].

\bibitem{Fan:2013yva}
J.~Fan, A.~Katz, L.~Randall and M.~Reece, \emph{{Double-Disk Dark Matter}},
  \href{http://dx.doi.org/10.1016/j.dark.2013.07.001}{\emph{Phys. Dark Univ.}
  {\bf 2} (2013) 139--156}, [\href{http://arxiv.org/abs/1303.1521}{{\tt
  1303.1521}}].

\bibitem{Foot:2014uba}
R.~Foot and S.~Vagnozzi, \emph{{Dissipative hidden sector dark matter}},
  \href{http://dx.doi.org/10.1103/PhysRevD.91.023512}{\emph{Phys. Rev.} {\bf
  D91} (2015) 023512}, [\href{http://arxiv.org/abs/1409.7174}{{\tt
  1409.7174}}].

\bibitem{Chacko:2005pe}
Z.~Chacko, H.-S. Goh and R.~Harnik, \emph{{The Twin Higgs: Natural electroweak
  breaking from mirror symmetry}},
  \href{http://dx.doi.org/10.1103/PhysRevLett.96.231802}{\emph{Phys. Rev.
  Lett.} {\bf 96} (2006) 231802},
  [\href{http://arxiv.org/abs/hep-ph/0506256}{{\tt hep-ph/0506256}}].

\bibitem{Chacko:2016hvu}
Z.~Chacko, N.~Craig, P.~J. Fox and R.~Harnik, \emph{{Cosmology in Mirror Twin
  Higgs and Neutrino Masses}}, {\emph{Submitted to: JHEP} (2016) },
  [\href{http://arxiv.org/abs/1611.07975}{{\tt 1611.07975}}].

\bibitem{Craig:2016lyx}
N.~Craig, S.~Koren and T.~Trott, \emph{{Cosmological Signals of a Mirror Twin
  Higgs}},  \href{http://arxiv.org/abs/1611.07977}{{\tt 1611.07977}}.

\bibitem{Jeong:2013eza}
K.~S. Jeong and F.~Takahashi, \emph{{Self-interacting Dark Radiation}},
  \href{http://dx.doi.org/10.1016/j.physletb.2013.07.001}{\emph{Phys. Lett.}
  {\bf B725} (2013) 134}, [\href{http://arxiv.org/abs/1305.6521}{{\tt
  1305.6521}}].

\bibitem{Buen-Abad:2015ova}
M.~A. Buen-Abad, G.~Marques-Tavares and M.~Schmaltz, \emph{{Non-Abelian dark
  matter and dark radiation}},
  \href{http://dx.doi.org/10.1103/PhysRevD.92.023531}{\emph{Phys. Rev.} {\bf
  D92} (2015) 023531}, [\href{http://arxiv.org/abs/1505.03542}{{\tt
  1505.03542}}].

\bibitem{Lesgourgues:2015wza}
J.~Lesgourgues, G.~Marques-Tavares and M.~Schmaltz, \emph{{Evidence for dark
  matter interactions in cosmological precision data?}},
  \href{http://dx.doi.org/10.1088/1475-7516/2016/02/037}{\emph{JCAP} {\bf 1602}
  (2016) 037}, [\href{http://arxiv.org/abs/1507.04351}{{\tt 1507.04351}}].

\bibitem{Chacko:2016kgg}
Z.~Chacko, Y.~Cui, S.~Hong, T.~Okui and Y.~Tsai, \emph{{Partially Acoustic Dark
  Matter, Interacting Dark Radiation, and Large Scale Structure}},
  \href{http://arxiv.org/abs/1609.03569}{{\tt 1609.03569}}.

\bibitem{Bashinsky:2003tk}
S.~Bashinsky and U.~Seljak, \emph{{Neutrino perturbations in CMB anisotropy and
  matter clustering}},
  \href{http://dx.doi.org/10.1103/PhysRevD.69.083002}{\emph{Phys. Rev.} {\bf
  D69} (2004) 083002}, [\href{http://arxiv.org/abs/astro-ph/0310198}{{\tt
  astro-ph/0310198}}].

\bibitem{Weinberg:2003ur}
S.~Weinberg, \emph{{Damping of tensor modes in cosmology}},
  \href{http://dx.doi.org/10.1103/PhysRevD.69.023503}{\emph{Phys. Rev.} {\bf
  D69} (2004) 023503}, [\href{http://arxiv.org/abs/astro-ph/0306304}{{\tt
  astro-ph/0306304}}].

\bibitem{Hu:1995en}
W.~Hu and N.~Sugiyama, \emph{{Small scale cosmological perturbations: An
  Analytic approach}}, \href{http://dx.doi.org/10.1086/177989}{\emph{Astrophys.
  J.} {\bf 471} (1996) 542--570},
  [\href{http://arxiv.org/abs/astro-ph/9510117}{{\tt astro-ph/9510117}}].

\bibitem{Follin:2015hya}
B.~Follin, L.~Knox, M.~Millea and Z.~Pan, \emph{{First Detection of the
  Acoustic Oscillation Phase Shift Expected from the Cosmic Neutrino
  Background}},
  \href{http://dx.doi.org/10.1103/PhysRevLett.115.091301}{\emph{Phys. Rev.
  Lett.} {\bf 115} (2015) 091301}, [\href{http://arxiv.org/abs/1503.07863}{{\tt
  1503.07863}}].

\bibitem{Baumann:2015rya}
D.~Baumann, D.~Green, J.~Meyers and B.~Wallisch, \emph{{Phases of New Physics
  in the CMB}},
  \href{http://dx.doi.org/10.1088/1475-7516/2016/01/007}{\emph{JCAP} {\bf 1601}
  (2016) 007}, [\href{http://arxiv.org/abs/1508.06342}{{\tt 1508.06342}}].

\bibitem{Blas:2011rf}
D.~Blas, J.~Lesgourgues and T.~Tram, \emph{{The Cosmic Linear Anisotropy
  Solving System (CLASS) II: Approximation schemes}},
  \href{http://dx.doi.org/10.1088/1475-7516/2011/07/034}{\emph{JCAP} {\bf 1107}
  (2011) 034}, [\href{http://arxiv.org/abs/1104.2933}{{\tt 1104.2933}}].

\bibitem{Audren:2012wb}
B.~Audren, J.~Lesgourgues, K.~Benabed and S.~Prunet, \emph{{Conservative
  Constraints on Early Cosmology: an illustration of the Monte Python
  cosmological parameter inference code}},
  \href{http://dx.doi.org/10.1088/1475-7516/2013/02/001}{\emph{JCAP} {\bf 1302}
  (2013) 001}, [\href{http://arxiv.org/abs/1210.7183}{{\tt 1210.7183}}].

\bibitem{Aghanim:2015xee}
{\scshape Planck} collaboration, N.~Aghanim et~al., \emph{{Planck 2015 results.
  XI. CMB power spectra, likelihoods, and robustness of parameters}},
  {\emph{Submitted to: Astron. Astrophys.} (2015) },
  [\href{http://arxiv.org/abs/1507.02704}{{\tt 1507.02704}}].

\bibitem{Adam:2016hgk}
{\scshape Planck} collaboration, R.~Adam et~al., \emph{{Planck intermediate
  results. XLVII. Planck constraints on reionization history}},
  \href{http://arxiv.org/abs/1605.03507}{{\tt 1605.03507}}.

\bibitem{Mangilli:2015xya}
A.~Mangilli, S.~Plaszczynski and M.~Tristram, \emph{{Large-scale cosmic
  microwave background temperature and polarization cross-spectra
  likelihoods}}, \href{http://dx.doi.org/10.1093/mnras/stv1733}{\emph{Mon. Not.
  Roy. Astron. Soc.} {\bf 453} (2015) 3174--3189},
  [\href{http://arxiv.org/abs/1503.01347}{{\tt 1503.01347}}].

\bibitem{Beutler:2011hx}
F.~Beutler, C.~Blake, M.~Colless, D.~H. Jones, L.~Staveley-Smith, L.~Campbell
  et~al., \emph{{The 6dF Galaxy Survey: Baryon Acoustic Oscillations and the
  Local Hubble Constant}},
  \href{http://dx.doi.org/10.1111/j.1365-2966.2011.19250.x}{\emph{Mon. Not.
  Roy. Astron. Soc.} {\bf 416} (2011) 3017--3032},
  [\href{http://arxiv.org/abs/1106.3366}{{\tt 1106.3366}}].

\bibitem{Ross:2014qpa}
A.~J. Ross, L.~Samushia, C.~Howlett, W.~J. Percival, A.~Burden and M.~Manera,
  \emph{{The clustering of the SDSS DR7 main Galaxy sample – I. A 4 per cent
  distance measure at $z = 0.15$}},
  \href{http://dx.doi.org/10.1093/mnras/stv154}{\emph{Mon. Not. Roy. Astron.
  Soc.} {\bf 449} (2015) 835--847}, [\href{http://arxiv.org/abs/1409.3242}{{\tt
  1409.3242}}].

\bibitem{Anderson:2013zyy}
{\scshape BOSS} collaboration, L.~Anderson et~al., \emph{{The clustering of
  galaxies in the SDSS-III Baryon Oscillation Spectroscopic Survey: baryon
  acoustic oscillations in the Data Releases 10 and 11 Galaxy samples}},
  \href{http://dx.doi.org/10.1093/mnras/stu523}{\emph{Mon. Not. Roy. Astron.
  Soc.} {\bf 441} (2014) 24--62}, [\href{http://arxiv.org/abs/1312.4877}{{\tt
  1312.4877}}].

\bibitem{Riess:2016jrr}
A.~G. Riess et~al., \emph{{A 2.4\% Determination of the Local Value of the
  Hubble Constant}},
  \href{http://dx.doi.org/10.3847/0004-637X/826/1/56}{\emph{Astrophys. J.} {\bf
  826} (2016) 56}, [\href{http://arxiv.org/abs/1604.01424}{{\tt 1604.01424}}].

\bibitem{Heymans:2013fya}
C.~Heymans et~al., \emph{{CFHTLenS tomographic weak lensing cosmological
  parameter constraints: Mitigating the impact of intrinsic galaxy
  alignments}}, \href{http://dx.doi.org/10.1093/mnras/stt601}{\emph{Mon. Not.
  Roy. Astron. Soc.} {\bf 432} (2013) 2433},
  [\href{http://arxiv.org/abs/1303.1808}{{\tt 1303.1808}}].

\bibitem{Ade:2013lmv}
{\scshape Planck} collaboration, P.~A.~R. Ade et~al., \emph{{Planck 2013
  results. XX. Cosmology from Sunyaev–Zeldovich cluster counts}},
  \href{http://dx.doi.org/10.1051/0004-6361/201321521}{\emph{Astron.
  Astrophys.} {\bf 571} (2014) A20},
  [\href{http://arxiv.org/abs/1303.5080}{{\tt 1303.5080}}].

\bibitem{Henderson:2015nzj}
S.~W. Henderson et~al., \emph{{Advanced ACTPol Cryogenic Detector Arrays and
  Readout}}, \href{http://dx.doi.org/10.1007/s10909-016-1575-z}{\emph{J. Low.
  Temp. Phys.} {\bf 184} (2016) 772--779},
  [\href{http://arxiv.org/abs/1510.02809}{{\tt 1510.02809}}].

\bibitem{Benson:2014qhw}
{\scshape SPT-3G} collaboration, B.~A. Benson et~al., \emph{{SPT-3G: A
  Next-Generation Cosmic Microwave Background Polarization Experiment on the
  South Pole Telescope}},
  \href{http://dx.doi.org/10.1117/12.2057305}{\emph{Proc. SPIE Int. Soc. Opt.
  Eng.} {\bf 9153} (2014) 91531P}, [\href{http://arxiv.org/abs/1407.2973}{{\tt
  1407.2973}}].

\bibitem{Suzuki:2015zzg}
{\scshape POLARBEAR} collaboration, A.~Suzuki et~al., \emph{{The POLARBEAR-2
  and the Simons Array Experiment}},
  \href{http://dx.doi.org/10.1007/s10909-015-1425-4}{\emph{J. Low. Temp. Phys.}
  {\bf 184} (2016) 805--810}, [\href{http://arxiv.org/abs/1512.07299}{{\tt
  1512.07299}}].

\bibitem{Laureijs:2011gra}
{\scshape EUCLID} collaboration, R.~Laureijs et~al., \emph{{Euclid Definition
  Study Report}},  \href{http://arxiv.org/abs/1110.3193}{{\tt 1110.3193}}.

\bibitem{Caruana:2013qua}
J.~Caruana, A.~J. Bunker, S.~M. Wilkins, E.~R. Stanway, S.~Lorenzoni, M.~J.
  Jarvis et~al., \emph{{Spectroscopy of $z \sim$ 7 candidate galaxies: Using
  Lyman $\alpha$ to constrain the neutral fraction of hydrogen in the
  high-redshift universe}},
  \href{http://dx.doi.org/10.1093/mnras/stu1341}{\emph{Mon. Not. Roy. Astron.
  Soc.} {\bf 443} (2014) 2831--2842},
  [\href{http://arxiv.org/abs/1311.0057}{{\tt 1311.0057}}].

\bibitem{Cyr-Racine:2013fsa}
F.-Y. Cyr-Racine, R.~de~Putter, A.~Raccanelli and K.~Sigurdson,
  \emph{{Constraints on Large-Scale Dark Acoustic Oscillations from
  Cosmology}}, \href{http://dx.doi.org/10.1103/PhysRevD.89.063517}{\emph{Phys.
  Rev.} {\bf D89} (2014) 063517}, [\href{http://arxiv.org/abs/1310.3278}{{\tt
  1310.3278}}].

\bibitem{Riess:2011yx}
A.~G. Riess et~al., \emph{{A 3\% Solution: Determination of the Hubble Constant
  with the Hubble Space Telescope and Wide Field Camera 3}},
  \href{http://dx.doi.org/10.1088/0004-637X/732/2/129,
  10.1088/0004-637X/730/2/119}{\emph{Astrophys. J.} {\bf 730} (2011) 119},
  [\href{http://arxiv.org/abs/1103.2976}{{\tt 1103.2976}}].

\bibitem{2013ApJ...766...70S}
S.~H. Suyu et~al., \emph{{Two Accurate Time-delay Distances from Strong
  Lensing: Implications for Cosmology}},
  \href{http://dx.doi.org/10.1088/0004-637X/766/2/70}{\emph{APJ} {\bf 766}
  (Apr., 2013) 70}, [\href{http://arxiv.org/abs/1208.6010}{{\tt 1208.6010}}].

\bibitem{Bernal:2016gxb}
J.~L. Bernal, L.~Verde and A.~G. Riess, \emph{{The trouble with $H_0$}},
  \href{http://dx.doi.org/10.1088/1475-7516/2016/10/019}{\emph{JCAP} {\bf 1610}
  (2016) 019}, [\href{http://arxiv.org/abs/1607.05617}{{\tt 1607.05617}}].

\bibitem{MacCrann:2014wfa}
N.~MacCrann, J.~Zuntz, S.~Bridle, B.~Jain and M.~R. Becker, \emph{{Cosmic
  Discordance: Are Planck CMB and CFHTLenS weak lensing measurements out of
  tune?}}, \href{http://dx.doi.org/10.1093/mnras/stv1154}{\emph{Mon. Not. Roy.
  Astron. Soc.} {\bf 451} (2015) 2877--2888},
  [\href{http://arxiv.org/abs/1408.4742}{{\tt 1408.4742}}].

\bibitem{Fu:2014loa}
L.~Fu et~al., \emph{{CFHTLenS: Cosmological constraints from a combination of
  cosmic shear two-point and three-point correlations}},
  \href{http://dx.doi.org/10.1093/mnras/stu754}{\emph{Mon. Not. Roy. Astron.
  Soc.} {\bf 441} (2014) 2725--2743},
  [\href{http://arxiv.org/abs/1404.5469}{{\tt 1404.5469}}].

\bibitem{Ade:2015zua}
{\scshape Planck} collaboration, P.~A.~R. Ade et~al., \emph{{Planck 2015
  results. XV. Gravitational lensing}},
  \href{http://arxiv.org/abs/1502.01591}{{\tt 1502.01591}}.

\bibitem{2012ApJ...755...70R}
{Reichardt, C.~L. et al.}, \emph{{A Measurement of Secondary Cosmic Microwave
  Background Anisotropies with Two Years of South Pole Telescope
  Observations}},
  \href{http://dx.doi.org/10.1088/0004-637X/755/1/70}{\emph{APJ} {\bf 755}
  (Aug., 2012) 70}, [\href{http://arxiv.org/abs/1111.0932}{{\tt 1111.0932}}].

\bibitem{Hasselfield:2013wf}
M.~Hasselfield et~al., \emph{{The Atacama Cosmology Telescope:
  Sunyaev-Zel'dovich selected galaxyclusters at 148 GHz from three seasons of
  data}}, \href{http://dx.doi.org/10.1088/1475-7516/2013/07/008}{\emph{JCAP}
  {\bf 1307} (2013) 008}, [\href{http://arxiv.org/abs/1301.0816}{{\tt
  1301.0816}}].

\bibitem{Ade:2015fva}
{\scshape Planck} collaboration, P.~A.~R. Ade et~al., \emph{{Planck 2015
  results. XXIV. Cosmology from Sunyaev-Zeldovich cluster counts}},
  \href{http://arxiv.org/abs/1502.01597}{{\tt 1502.01597}}.

\end{thebibliography}\endgroup
\bibliographystyle{JHEP}

\end{document}